%% file: KM3NeT_PUB_2023_006-ARCA230.tex
\documentclass[pdftex,twocolumn,epjc3]{svjour3}          

\RequirePackage[T1]{fontenc}

\smartqed  

\RequirePackage{graphicx}
\RequirePackage{mathptmx}      
\RequirePackage{flushend}
\RequirePackage[numbers,sort&compress]{natbib}
\RequirePackage[colorlinks,citecolor=blue,urlcolor=blue,linkcolor=blue]{hyperref}

\usepackage{graphicx}%
\usepackage{multirow}%
\usepackage{amsmath,amssymb,amsfonts}%
\usepackage{mathrsfs}%
\usepackage[title]{appendix}%
\usepackage{xcolor}%
\usepackage{textcomp}%
\usepackage{manyfoot}%
\usepackage{booktabs}%
\usepackage{algorithm}%
\usepackage{algorithmicx}%
\usepackage{algpseudocode}%
\usepackage{listings}%
\usepackage{subfig}
\usepackage{hyperref}

\journalname{Eur. Phys. J. C}

\begin{document}

\title{Astronomy potential of KM3NeT/ARCA}

\author{
S.~Aiello\thanksref{a} \and
A.~Albert\thanksref{b,bc} \and
M.~Alshamsi\thanksref{d,c} \and
S. Alves Garre\thanksref{e} \and
Z.~Aly\thanksref{c} \and
A. Ambrosone\thanksref{g,f} \and
F.~Ameli\thanksref{h} \and
M.~Andre\thanksref{i} \and
E.~Androutsou\thanksref{j} \and
M.~Anguita\thanksref{k}
% ...
\\(KM3NeT Collaboration\thanksref{emailkm3}
}

\thankstext{e1}{e-mail: thijsvaneeden@gmail.com}
\thankstext{emailkm3}{e-mail: km3net-pc@km3net.de}

\institute{
\setlength{\parindent}{0pt}
INFN, Sezione di Catania, (INFN-CT) Via Santa Sofia 64, Catania, 95123 ~Italy\label{a} \and
Universit{\'e}~de~Strasbourg,~CNRS,~IPHC~UMR~7178,~F-67000~Strasbourg,~France\label{b} \and
Aix~Marseille~Univ,~CNRS/IN2P3,~CPPM,~Marseille,~France\label{c} \and
University of Sharjah, Sharjah Academy for Astronomy, Space Sciences, and Technology, University Campus - POB 27272, Sharjah, - United Arab Emirates\label{d} \and
IFIC - Instituto de F{\'\i}sica Corpuscular (CSIC - Universitat de Val{\`e}ncia), c/Catedr{\'a}tico Jos{\'e} Beltr{\'a}n, 2, 46980 Paterna, Valencia, Spain\label{e} \and
INFN, Sezione di Napoli, Complesso Universitario di Monte S. Angelo, Via Cintia ed. G, Napoli, 80126 Italy\label{f} \and
Universit{\`a} di Napoli ``Federico II'', Dip. Scienze Fisiche ``E. Pancini'', Complesso Universitario di Monte S. Angelo, Via Cintia ed. G, Napoli, 80126 Italy\label{g} \and
INFN, Sezione di Roma, Piazzale Aldo Moro 2, Roma, 00185 Italy\label{h} \and
Universitat Polit{\`e}cnica de Catalunya, Laboratori d'Aplicacions Bioac{\'u}stiques, Centre Tecnol{\`o}gic de Vilanova i la Geltr{\'u}, Avda. Rambla Exposici{\'o}, s/n, Vilanova i la Geltr{\'u}, 08800 Spain\label{i} \and
NCSR Demokritos, Institute of Nuclear and Particle Physics, Ag. Paraskevi Attikis, Athens, 15310 Greece\label{j} \and
University of Granada, Dept.~of Computer Architecture and Technology/CITIC, 18071 Granada, Spain\label{k}
% ...
\and
Universit{\'e} de Haute Alsace, rue des Fr{\`e}res Lumi{\`e}re, 68093 Mulhouse Cedex, France\label{bc}
% ...
}

\date{Received: \today / Accepted: T.B.D.}

\onecolumn

\maketitle

\twocolumn    

\begin{abstract}
The KM3NeT/ARCA neutrino detector is currently under construction at 3500 m depth offshore Capo Passero, Sicily, in the Mediterranean Sea. The main science objectives are the detection of high-energy cosmic neutrinos and the discovery of their sources. Simulations were conducted for the full KM3NeT/ARCA detector, instrumenting a volume of 1 km$^3$, to estimate the sensitivity and discovery potential to point-like neutrino sources. This paper covers the reconstruction of track- and shower-like signatures, as well as the criteria employed for neutrino event selection. With an angular resolution below 0.1$^\circ$ for tracks and under 2$^\circ$ for showers, the sensitivity to point-like neutrino sources surpasses existing observed limits across the entire sky.
\end{abstract}

\input{introduction}

\input{signatures}

\input{detector}

\input{simulation}

\input{reconstruction}

\newpage

\input{selection}

\newpage

\input{point_sources}

\input{systematics}

\input{conclusions}

\section{Acknowledgements} The authors acknowledge the financial support of the funding agencies:
Czech Science Foundation (GA\u CR 24-12702S);
Agence Nationale de la Recherche (contract ANR-15-CE31-0020), Centre National de la Recherche Scientifique (CNRS), Commission Europ\'eenne (FEDER fund and Marie Curie Program), LabEx UnivEarthS (ANR-10-LABX-0023 and ANR-18-IDEX-0001), Paris \^Ile-de-France Region, France;
Shota Rustaveli National Science Foundation of Georgia (SRNSFG, FR-22-13708), Georgia;
The General Secretariat of Research and Innovation (GSRI), Greece;
Istituto Nazionale di Fisica Nucleare (INFN) and Ministero dell’Universit{\`a} e della Ricerca (MUR), through PRIN 2022 program (Grant PANTHEON 2022E2J4RK, Next Generation EU) and PON R\&I program (Avviso n. 424 del 28 febbraio 2018, Progetto PACK-PIR01 00021), Italy; A. De Benedittis, R. Del Burgo, W. Idrissi Ibnsalih, A. Nayerhoda, G. Papalashvili, I. C. Rea, S. Santanastaso, A. Simonelli have been supported by the Italian Ministero dell'Universit{\`a} e della Ricerca (MUR), Progetto CIR01 00021 (Avviso n. 2595 del 24 dicembre 2019);
Ministry of Higher Education, Scientific Research and Innovation, Morocco, and the Arab Fund for Economic and Social Development, Kuwait;
Nederlandse organisatie voor Wetenschappelijk Onderzoek (NWO), the Netherlands;
The National Science Centre, Poland (2021/41/N/ST2/01177); The grant “AstroCeNT: Particle Astrophysics Science and Technology Centre”, carried out within the International Research Agendas programme of the Foundation for Polish Science financed by the European Union under the European Regional Development Fund;
National Authority for Scientific Research (ANCS), Romania;
Slovak Research and Development Agency under Contract No. APVV-22-0413; Ministry of Education, Research, Development and Youth of the Slovak Republic;
MCIN for PID2021-124591NB-C41, -C42, -C43, funded by MCIN/AEI/10.13039/501100011033 and by "ERDF A way of making Europe", for ASFAE/2022/014, ASFAE/2022 /023, with funding from the EU NextGenerationEU (PRTR-C17.I01), Generalitat Valenciana, and for CSIC-INFRA23013, Generalitat Valenciana for PROMETEO /2020/019, for Grant AST22\_6.2 with funding from Consejer\'{\i}a de Universidad, Investigaci\'on e Innovaci\'on and Gobierno de Espa\~na and European Union - NextGenerationEU, for CIDEGENT/2018/034, /2019/043, /2020/049, /2021/23 and for GRISOLIAP/2021/192 and EU for MSC/101025085, Spain;
The European Union's Horizon 2020 Research and Innovation Programme (ChETEC-INFRA - Project no. \newline 101008324).

\bibliographystyle{unsrt}       
\bibliography{biblio}

\end{document}

%% file: introduction.tex
\section{Introduction}

The discovery of the diffuse high-energy cosmic neutrino flux by the IceCube Collaboration \cite{icecube2013evidence,abbasi2022improved,abbasi2021icecube} strengthened the field of neutrino astronomy. The first candidate neutrino source was identified thanks to the observation of a gamma-ray flare from the blazar TXS 0506+056 in temporal and spatial coincidence with a high-energy neutrino event detected in the IceCube detector \cite{icecube2018neutrino}. Recently, a significant excess of events from the direction of the nearby active galaxy NGC 1068 was observed \cite{icecube2022evidence}, and a diffuse neutrino emission from the Galactic plane was measured \cite{icecube2023observation}. KM3NeT is a research infrastructure that hosts two deep seawater neutrino telescopes: ARCA (Astroparticle Research with Cosmics in the Abyss) and ORCA (Oscillation Research with Cosmics in the Abyss). They use the same technology and detection principle, though their size and primary scientific objectives are different.

The ARCA detector is currently under construction at a depth of 3500 m on the seabed of the Mediterranean Sea, offshore Capo Passero, Sicily, Italy \cite{adrian2016letter}. The detector consists of a grid of optical modules that detect Cherenkov radiation induced in the medium by charged secondary particles created in neutrino interactions. Its primary purpose is to detect neutrinos with energies beyond a TeV with the aim of doing neutrino astronomy. The ARCA field of view for upgoing neutrinos covers mainly the Southern sky, enabling the study of potential Galactic sources. Upgoing neutrinos are promising in neutrino astronomy due to the shielding of the Earth against muons from air showers in the atmosphere induced by cosmic rays. The first scientific results from ARCA, obtained with partial detector configurations, have been presented in References \cite{muller2023point,vasilis2023diffuse,francesco2023galactic}. These analyses included up to 21 active detection units, out of the total of 230 that are planned. Once fully deployed, these units will collectively instrument a total volume of 1 km$^3$. The ORCA detector, under construction at a depth of 2500 m off the coast of Toulon, France, is specifically designed for low-energy neutrino detection and aims to determine the mass ordering of neutrinos through the study of oscillations of atmospheric neutrinos. ORCA, although not discussed in this article, also serves to broaden the scope of astronomical analyses to lower energies.

This article explores the capabilities of the complete \\ ARCA detector in detecting point-like neutrino sources using detailed Monte Carlo simulations. Improvements in the event reconstruction performance, coupled with the combination of the track and shower channels, result in improved sensitivities compared to those outlined in References \cite{adrian2016letter,aiello2019sensitivity}. The reconstruction algorithms and corresponding performances for different detection signatures are discussed, followed by a description of the neutrino selection criteria. Statistical analyses utilising a binned likelihood approach have been conducted to evaluate the sensitivity and discovery potential to various neutrino sources.

%% file: signatures.tex
\section{Detection signatures}

Neutrinos can only be detected through their weak interactions. At energies above several GeV, the dominant process is deep inelastic scattering, where the neutrino scatters off the quarks inside a nucleon. This process is mediated by either a $W^\pm$ boson in charged current (CC) interactions or a $Z^0$ boson in neutral current (NC) interactions. 

In the final state of both types of interactions high-energy hadrons generate a shower of particles. In the case of the exchange of a $W^\pm$ boson, a charged lepton is also produced, with a flavour that corresponds to the neutrino flavour eigenstate. Neutrino telescopes such as ARCA take advantage of the emission pattern of the Cherenkov light induced by the products of the neutrino interactions in the medium, in order to reconstruct the neutrino direction and energy. This results in two main detection signatures:
\begin{itemize}
    \item Track: muons from $\nu_\mu$ CC interactions and $\nu_\tau$ CC interactions where the $\tau$ decays into a muon (branching ratio $\sim 17\%$),
    \item Shower: hadronic showers from all-flavour NC interactions, electromagnetic showers from $\nu_e$ CC interactions and $\nu_\tau$ CC interactions where the $\tau$ decays either into an electron or into hadrons (branching ratio $\sim 83\%$).
\end{itemize}

Muons with TeV energies or higher can travel kilometers in seawater and can be detected even if the neutrino interaction happens far away from the detector. The best angular resolution is achieved for track-like events thanks to the long distances muons travel in the detector. Shower-like signatures are characterised by the deposition of their energy within tens of meters from the neutrino interaction vertex. This results in a worse angular resolution but improved energy resolution because the event can be contained in the detector.

%% file: detector.tex
\section{The KM3NeT/ARCA detector}

\subsection{Detector layout}

The main detector component of the KM3NeT experiment is the digital optical module (DOM) \cite{aiello2022km3net}. The DOM consists of a 44 cm diameter glass sphere housing 31 3-inch photomultiplier tubes (PMTs). The PMTs are surrounded by reflector rings that increase the acceptance by 20-30\%. Each DOM is equipped with data acquisition electronics and a piezo sensor, a compass and a tiltmeter for calibration purposes.

\begin{figure*}
\centering
\includegraphics[width=0.95\textwidth]{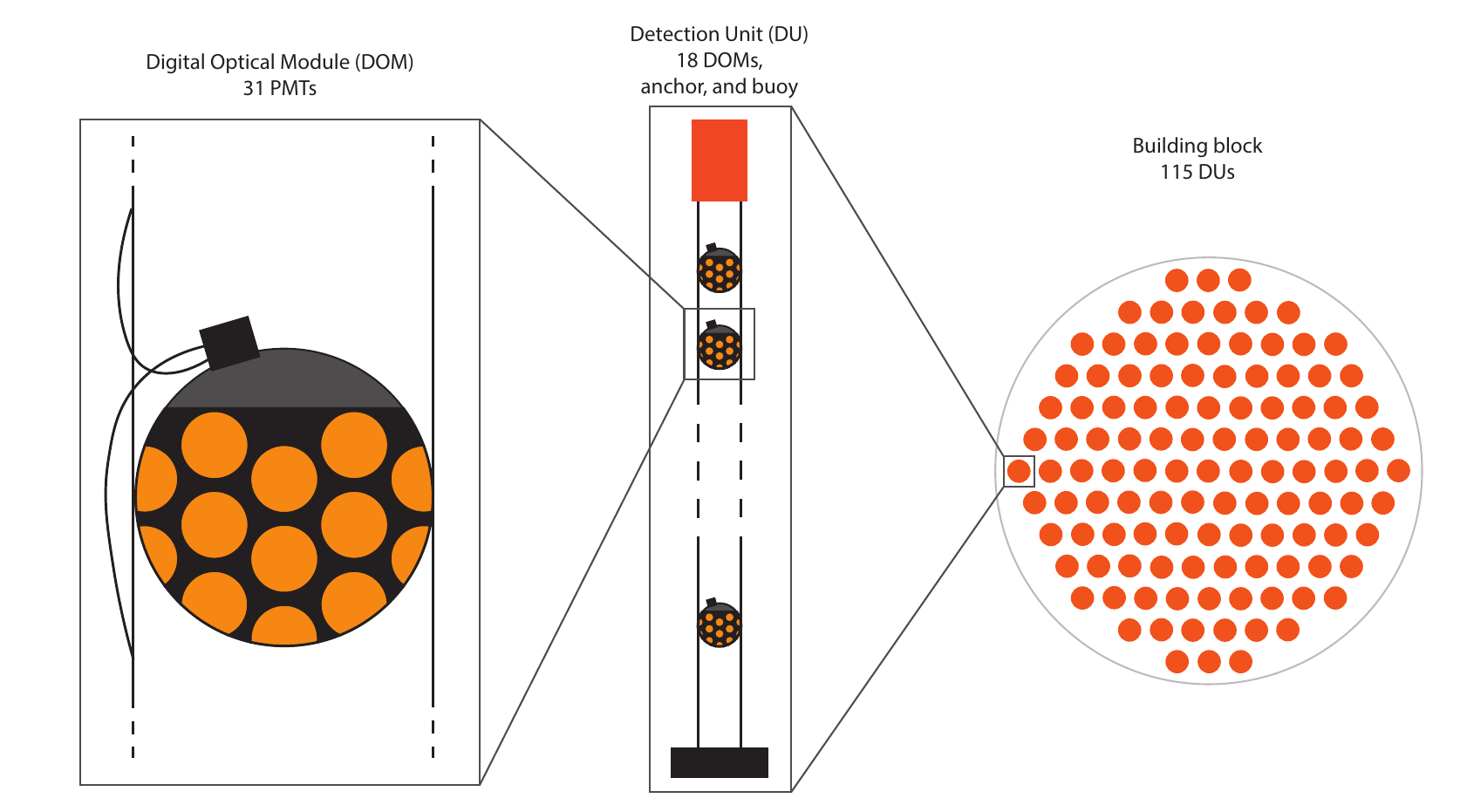}
\caption{The KM3NeT detector layout. The full ARCA detector will have 2 building blocks of 115 detection units each.}
\label{fig:detlayout}
\end{figure*}

The DOMs are attached to detection units that are anchored at the seabed. Each detection unit powers 18 DOMs and transports the data via optical fibers. The detection unit is kept vertical due to the buoyancy of the DOMs and the buoy attached to the top. The average vertical distance between DOMs along a detection unit is 36~m and the detection units are placed on the seafloor with an average spacing of 95~m. The spacing between the detector components of the ORCA detector is smaller in order to target lower neutrino energies. The detection units are grouped in building blocks, each comprising 115 detection units. A schematic view of this configuration is shown in Figure \ref{fig:detlayout}. The ARCA detector will consist of two building blocks. As of September 2023, a total of 28 detection units have been deployed.

\subsection{Optical background sources}

The most substantial background in neutrino telescopes originates from cosmic rays interacting with the atmosphere and producing air showers. Many of the muons coming from the decay of short-lived particles in the air shower reach the surface of the Earth, and high-energy muons can even reach the detector depth of 3.5 km. One of the first KM3NeT data analyses shows the depth dependence of the muon rate in Reference \cite{ageron2020dependence}. Other sources of optical background arise from bioluminescence and radioactive decays in the seawater. 

\subsection{Data acquisition}

The PMT signals are pulses that are digitised using a threshold discriminator. This reduces photon detection to a \textit{hit}, characterised by a hit time and a time-over-threshold. Data acquisition is based on an \textit{all-data-to-shore} concept. In this, all hits are transmitted to shore and processed in real time. Data are filtered using trigger algorithms which select subsets of hits, commonly referred to as an event, consistent with a relativistic particle crossing the detector. These hit patterns are distinguished from background through the use of time-position correlations. Random background hits are uncorrelated and hits from bioluminescent activity are correlated on a single optical module but uncorrelated between modules. All hits within a predefined time window around an event are written to disk.

\subsection{Calibration}

To fulfill the physics objectives of ARCA it is essential to conduct precise calibrations of the detector and acquire comprehensive knowledge about the environmental conditions affecting the detector operation. To achieve the envisaged pointing resolution below 0.1$^\circ$ for tracks, the positions and orientations of the DOMs should be measured with a resolution of 20 cm and 3$^\circ$ respectively and the time offsets of the PMTs with 1 ns accuracy. This work assumes that the required calibration accuracy is achieved. Calibration efforts using the operational detection units have achieved the desired timing accuracy \cite{louis2023time,agustin2023time} and measured the relative positions and orientations of the modules \cite{gatius2023dynamic}. The deficit of atmospheric muons due to the cosmic ray shadow from the Sun and the Moon was observed using the ORCA detector \cite{aiello2023first}. This analysis presented the first verification of the absolute pointing of the detector and is currently carried out for ARCA.

%% file: simulation.tex
\section{Simulation}

The general simulation framework used by KM3NeT exploits the past experience of the ANTARES detector \cite{albert2021monte}, though dedicated new software packages have been developed. The simulation chain for this analysis can be subdivided into several steps, briefly described hereafter.

\subsection{Event generation}

Neutrino interactions are simulated using the gSeaGen simulation framework \cite{aiello2020gseagen}. All interaction channels of neutrinos of all flavours are simulated over an energy range spanning from $10^2$ to $10^8$ GeV. Simulated neutrino events are then weighted according to different flux models for neutrinos of atmospheric and cosmic origin, in order to estimate their rate at the detector. The atmospheric neutrino flux consists of a conventional and a prompt contribution. The conventional contribution arises from the decay of light and long-lived mesons, such as charged pions and kaons, and dominates the flux composition up to 10-100 TeV. The prompt contribution comes from the decay of heavier and short-lived hadrons and is expected to have a harder energy spectrum than the conventional flux. In this work, the conventional component is represented by the 2006 flux model from \textit{Honda et al.} \cite{honda2007calculation}. The prompt component is taken from \textit{Enberg et al.} \cite{enberg2008prompt}. Both contributions are corrected to account for the primary cosmic ray knee according to the H3a composition model of cosmic rays \cite{gaisser2012spectrum}.
    
The flux of atmospheric muons is simulated with the MUPAGE software package \cite{carminati2008atmospheric}, which describes the rate of muons at a given depth in the sea or in ice by means of parametric formulas. The simulation includes both individual muons and multiple muons generated by the same cosmic ray air shower. As priority was given to simulate high-energy atmospheric muons, which may pass the event selections, the muon simulations reported in this paper have a lower-threshold bundle energy of 10 TeV. It was checked that lower-energy muons have no significant impact on the analysis presented here.

The atmospheric backgrounds are kept constant in the analysis, and the pseudo-experiments are drawn from these expectations using Poisson statistics.

\subsection{Particle propagation and light simulation}

The generated events are processed by an internal KM3NeT software package that simulates the particle propagation and the production of Cherenkov photons in water. The simulation of light from charged particles in the event is based on probability density functions (PDFs) describing the arrival time of photons on a PMT as a function of the energy of the particle, of the distance of a PMT from the particle trajectory, and of the incidence angle of light on the PMT \cite{de2023probability}. The description of light emitted along muon trajectories accounts for that of a minimum ionising particle, delta rays, and bremsstrahlung showers emitted along the muon track. Bremsstrahlung and pair production from electrons are simulated as an electromagnetic shower. Hadrons are assumed to produce similar light patterns as electromagnetic showers, but with a downscaled light yield. Dispersion, absorption, and scattering effects of photons are taken into account to describe the transmission of light. The arrival times of the simulated photo-electrons are stored after convolving the light yield from the processes described above with the angular acceptance and the quantum efficiency of the PMTs.

\subsection{Detector response}

The information on the detected photo-electrons is used as input for the software package dedicated to describe the response of the KM3NeT detector. This software simulates the PMT response to light, the front-end electronics, and random optical backgrounds. As a result, the simulated data stream is represented by a collection of hits that are in the same format as the data coming from the sea. The algorithm determines if an event is stored, by applying dedicated event triggering conditions - the same that are applied to data from the sea. These events are input to the reconstruction algorithms for tracks and showers described in the following section.

%% file: reconstruction.tex
\section{Event reconstruction}
\label{sec:reco}

\subsection{Track reconstruction}

\begin{figure*}
    \centering
    \includegraphics[width=0.9\textwidth]{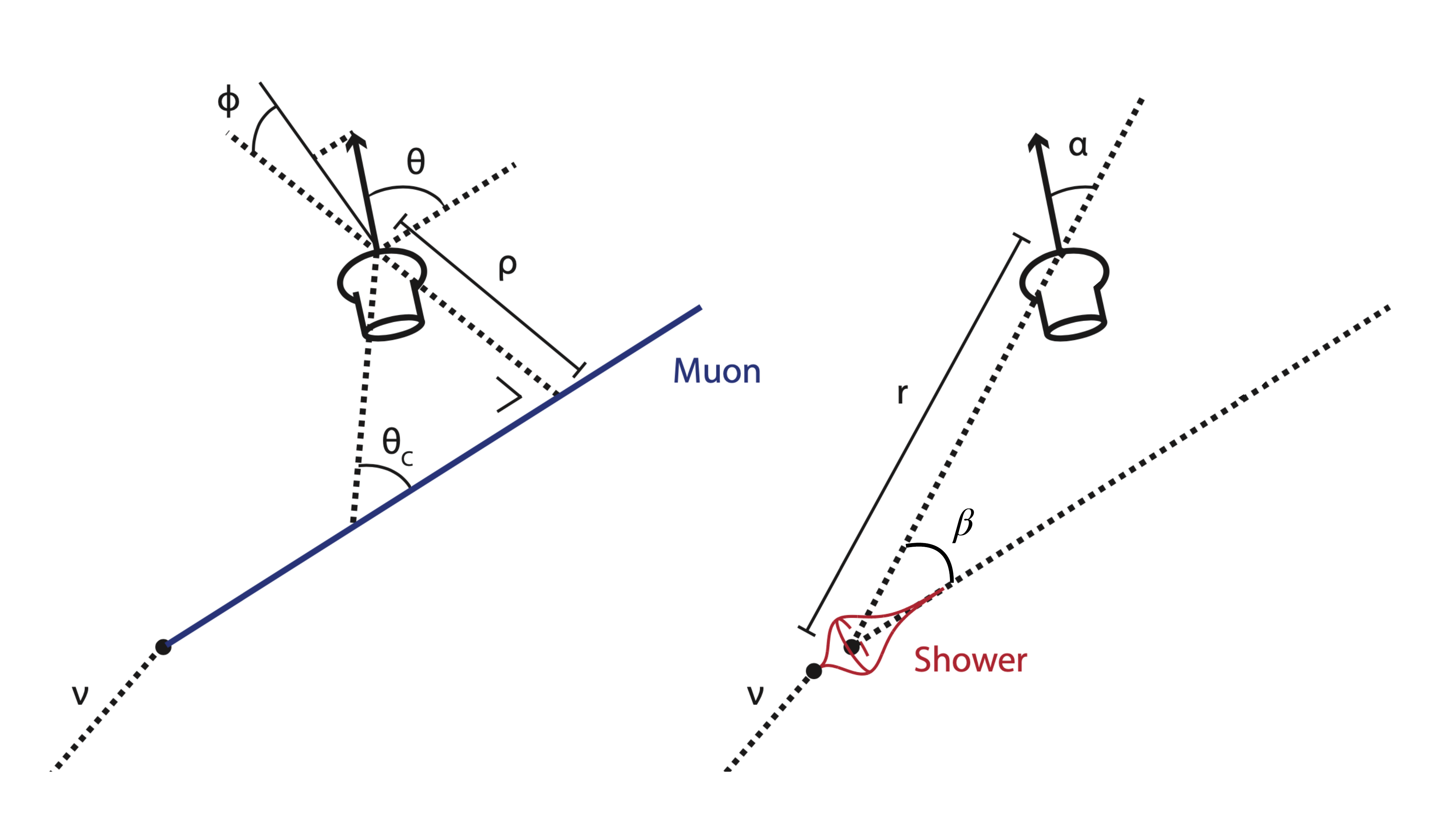}
    \caption{Configuration of a PMT with a muon track (left) and a shower (right) \cite{melis2021studying}.}
    \label{fig:recosketch}
\end{figure*}

Track reconstruction algorithms are used to determine the energy $E_\mu$, the direction $\vec{d}_\mu$ and the position $\vec{x}_\mu$ and time $t_\mu$ along the trajectory of a muon from the data \cite{melis2017km3net}. The PDF of the arrival time of Cherenkov light can be described by the following five parameters:
\begin{itemize}
    \item $\rho$: minimum distance between the muon trajectory and the PMT;
    \item $\theta, \phi$: PMT orientation angles with respect to the muon trajectory;
    \item $\Delta t$: difference between measured and expected hit time according to the Cherenkov hypothesis;
    \item $E_\mu$: energy of the muon.
\end{itemize}
A schematic of the configuration is shown in Figure \ref{fig:recosketch}.

In a first step, a predefined set of 20,000 directions covering the full sky is explored. In each direction, only the lateral position and the time along the trajectory of the muon need to be determined. By neglecting scattering and dispersion of light, this problem reduces to a linear fit. To maximise the probability that the right solution is maintained whilst limiting the CPU time, the best prefit solutions are passed to the next step. In the final two steps, the likelihood functions $\mathcal{L}(\vec{d}_\mu,\vec{x}_\mu,t_\mu)$ and $\mathcal{L}(E_\mu)$ are independently maximised. This maximisation uses all PMTs within a cylinder with a radius of 150 m around the previously found muon tracks.

\subsection{Shower reconstruction}

The shower reconstruction program follows a two-step procedure \cite{melis2017km3net}. In the first step, the position ($\vec{x}_s$) and time ($t_s$) of the shower maximum are estimated using coincident hits on the same optical module within 20 ns. The shower maximum refers to the point along the path of a high-energy particle shower where the number of secondary particles and light reaches its maximum value. The position and time are found by minimising the function
\begin{equation}
    M(\vec{x}_s,t_s) = \sum_{i \in \text{hits}} \sqrt{1+(t_i-\hat{t}_i)},
\end{equation}
where $t_i$ is the measured hit time and $\hat{t}_i$ the expected hit time assuming unscattered propagation of the photon. Particle showers emit light across several meters, yet their elongation is small with respect to the spacing of the instrument. For this reason, the expected hit time can be calculated assuming a spherical light pattern emitted from the vertex according to
\begin{equation}
    \hat{t}_i = t_s + \frac{d}{c_{\rm water}},
\end{equation}
where $d$ is the distance from the PMT to the assumed vertex and $c_{\rm water}$ is the speed of light in water. 

The final fit tests different direction hypotheses, where the estimated vertex position of the previous step acts as a pivot. Twelve starting directions are chosen isotropically over the sky where the following likelihood is maximised using the PMTs that registered a hit (\textit{hit} PMTs) and the PMTs that did not (\textit{no hit} PMTs):
\begin{align}
    & \log \mathcal{L} (E_{\rm shower},\vec{d}_{\rm shower}) = \\
    & \sum_{i \in \text{hit PMTs}} \log (P_i^{\text{hit}}) + \sum_{i \in \text{no hit PMTs}} \log (P_i^{\text{no hit}}), \notag \\
\end{align}
where
\begin{align}
    & P_i^{\text{hit}} = 1 - P_i^{\text{no hit}} = \\
    & 1 - \exp \Big( - \mu_{\text{sig}}(r_i,\beta_i,a_i,E_{\rm shower}) - R_{\rm bg} \cdot T \Big) \notag.
\end{align}
The expected number of signal events $\mu_{\text{sig}}$ is obtained from a PDF based on Monte Carlo simulations. $R_{\rm bg}$ is the expected background rate in the time window $T$. The PDF depends on the following parameters:
\begin{itemize}
    \item $r$: the distance from the vertex to the PMT;
    \item $\beta$: the angle between the neutrino direction and the vector between the vertex and the PMT;
    \item $a$: the angle between the normal vector of the PMT and the vector between the vertex and the PMT;
    \item $E_{\rm shower}$: the energy of the shower;
\end{itemize}
A schematic of the configuration is shown in Figure \ref{fig:recosketch}.

%% file: selection.tex
\section{Neutrino selections}
\label{sec:selection}

Most of the events collected in a neutrino telescope are due to atmospheric muons and neutrinos which represent the main sources of background. The analysis in this paper replicates the same situation using Monte Carlo simulations. This allows to define a set of selection criteria providing event samples with a known level of background contamination. Selection requirements are defined separately for each of the two observation channels: track and shower. The events surviving a primary selection are then used to train a Boosted Decision Tree (BDT) model using the TMVA software package \cite{voss2009tmva}, whose output allows to determine the final selections. The selections are finalised to get a high neutrino purity, defined as the fraction of neutrinos of atmospheric and cosmic origin in the final sample. The expected rates of cosmic neutrino events are obtained using a neutrino flux of
\begin{equation}
    \Phi^{\nu_i + \bar{\nu}_i} = 1.2 \times 10^{-8} \left( \frac{E_{\nu}}{\rm GeV} \right)^{-2} \text{ GeV}^{-1} \rm cm^{-2} s^{-1} sr^{-1}
\label{eq:cosflux}
\end{equation}
where $i=e,\mu,\tau$.

\subsection{Track selection}

Upgoing and horizontal events with reconstructed zenith angle $\theta > 80^\circ$ are selected. Vertical downgoing tracks correspond to $\theta=0^\circ$, making this cut effective in excluding downgoing atmospheric muons. A residual contamination of atmospheric muons is due to misreconstructed events and can be further suppressed with a BDT. Input variables for the BDT training include, listed in order of their importance in the discrimination power between signal and background: the estimated number of photo-electrons along the muon trajectory, the estimated error on the reconstructed direction of the track, the fitted track length, the fit quality of the track reconstruction and the reconstructed muon energy.

\begin{figure*}
    \centering
    \subfloat[\centering  ]{{\includegraphics[width=0.45\textwidth]{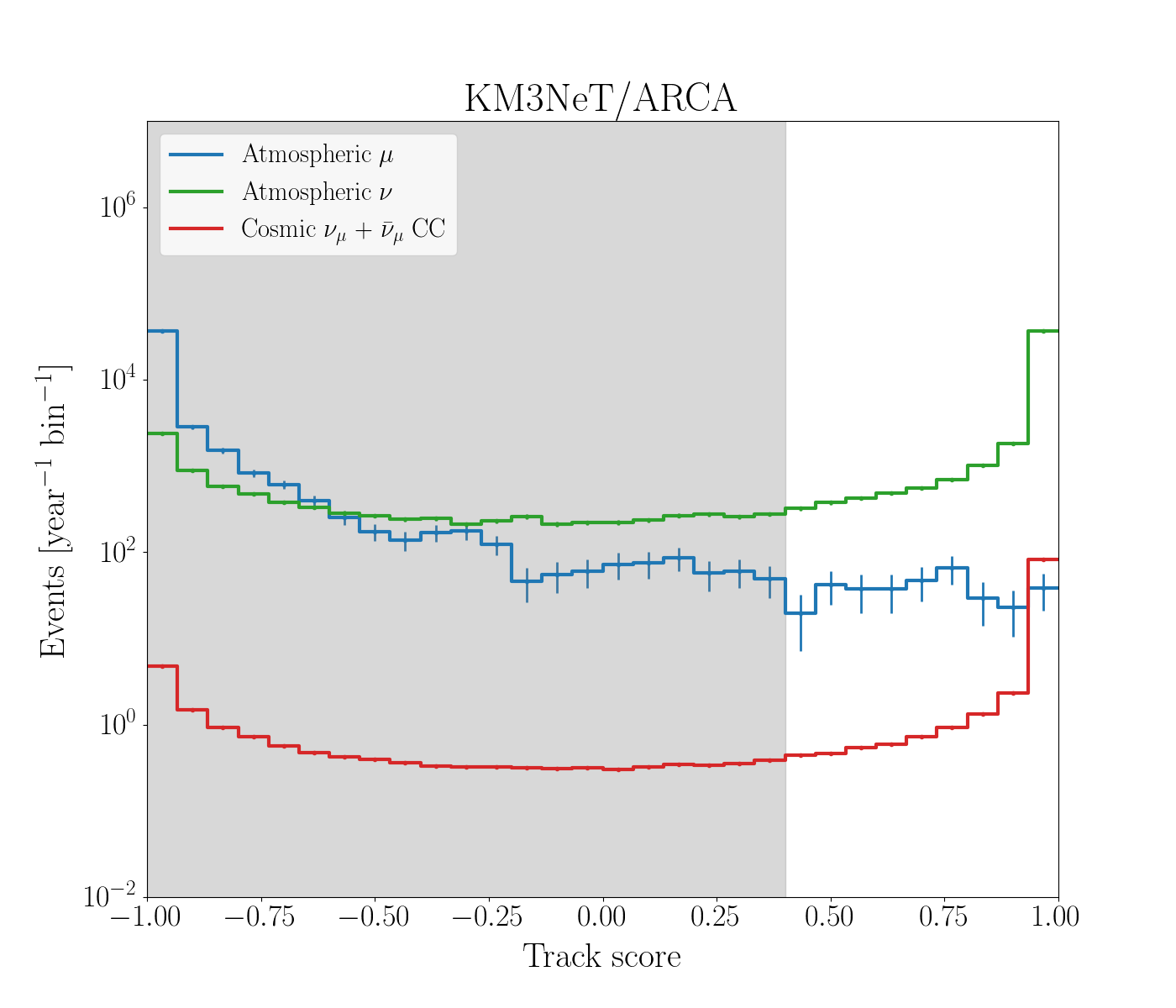} }}%
    \qquad
    \subfloat[\centering  ]{{\includegraphics[width=0.45\textwidth]{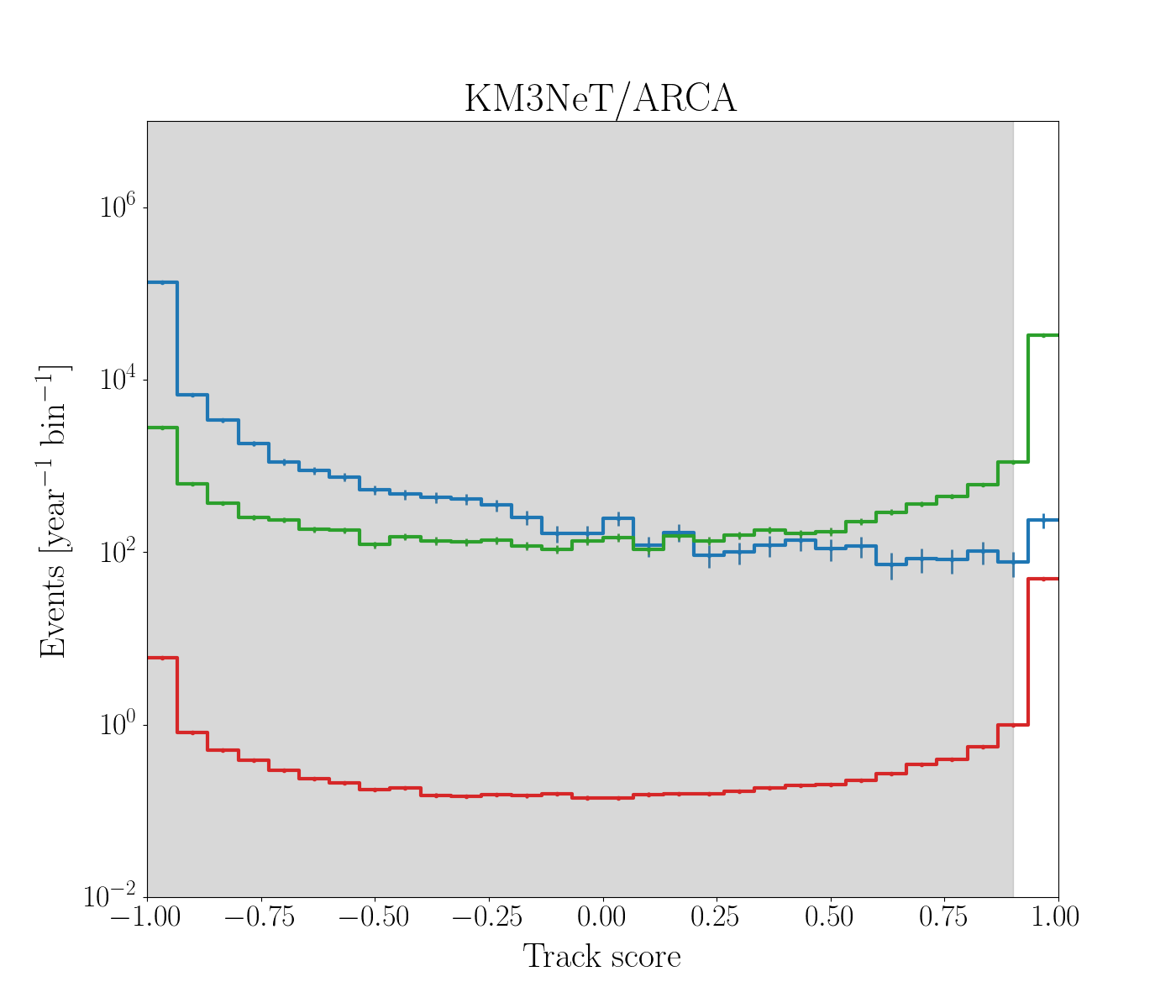}}}%
    \caption{ARCA track classifier score for $\theta > 100^\circ$ (a) and $80^\circ < \theta < 100^\circ$ (b). The shaded region covers events that are rejected. The expected cosmic event rate is obtained using the flux from equation \ref{eq:cosflux}.}
    \label{fig:trackbdt}%
\end{figure*}

The BDT provides a \textit{track score} where higher values indicate a higher probability that the event was induced by a neutrino interaction. Distributions for the track score are shown in Figure \ref{fig:trackbdt} for upgoing ($\theta > 100^\circ$) and horizontal events ($80^\circ < \theta < 100^\circ$), where the shaded region covers events that are rejected.

\begin{figure*}
    \centering
    \subfloat[\centering  ]{{\includegraphics[width=0.45\textwidth]{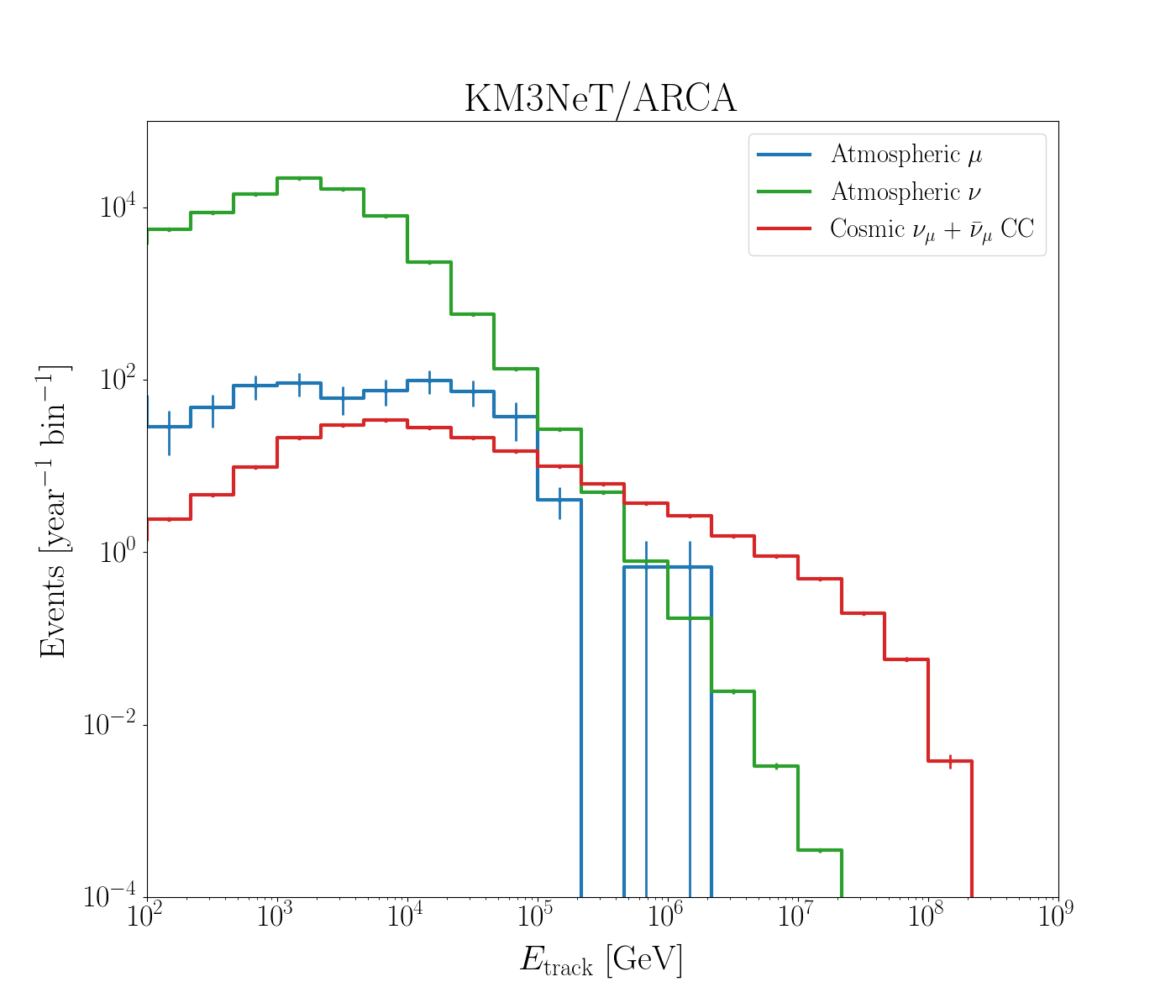}}}%
    \qquad
    \subfloat[\centering  ]{{\includegraphics[width=0.45\textwidth]{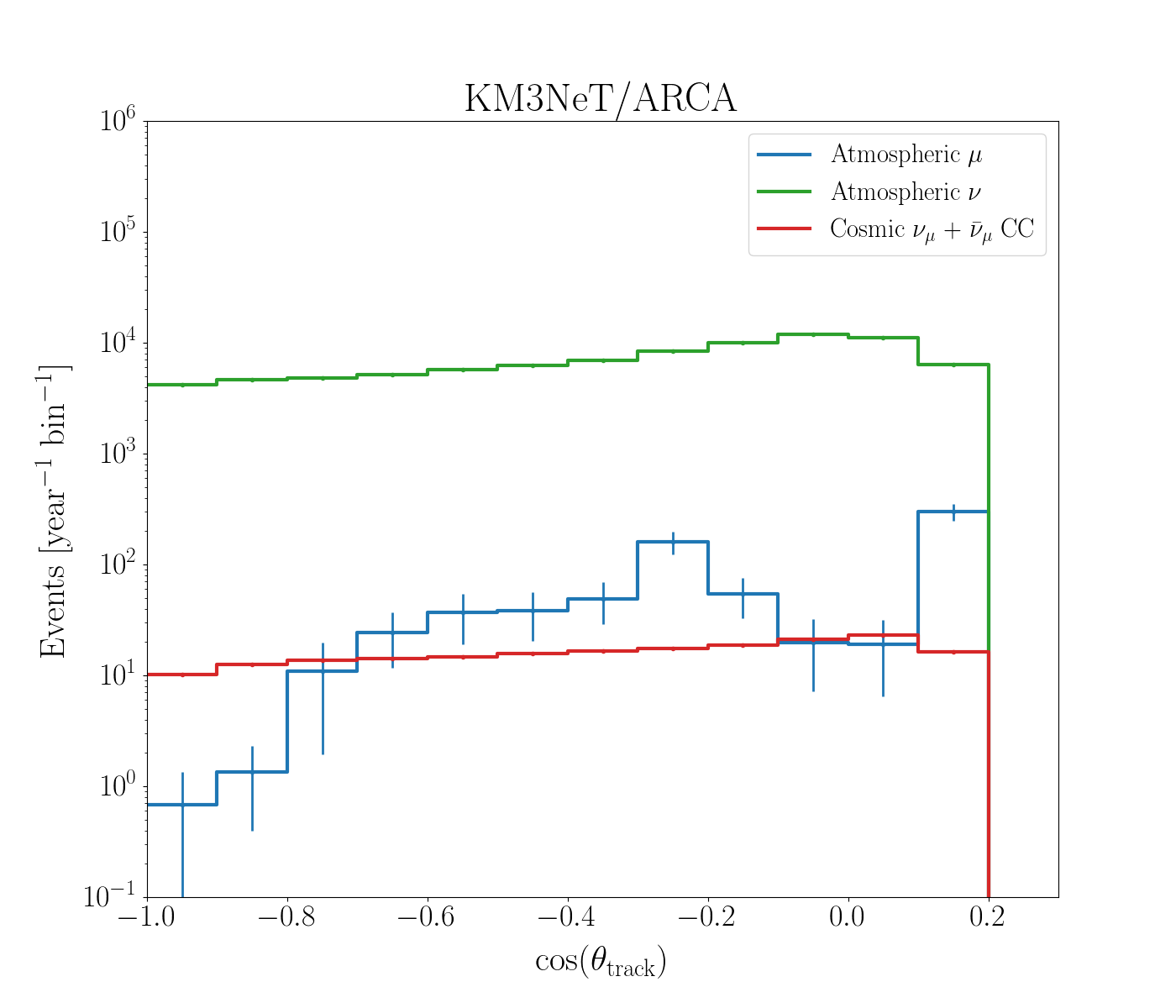} }}%
    \caption{ARCA event distribution as a function of the reconstructed energy (a) and zenith angle (b) of events passing the BDT track score requirement. The expected cosmic event rate is obtained using the flux from equation \ref{eq:cosflux}.}
    \label{fig:trackdistributions}%
\end{figure*}

\begin{table*}
\centering
\begin{tabular}{llll}
\hline
  &  Trigger [yr$^{-1}$] &  Zenith [yr$^{-1}$] & BDT [yr$^{-1}$] \\
\hline
Atmospheric $\mu$ (> 10 TeV) &  $8.1 \times 10^7$ &  $5.3 \times 10^5$ &  710 \\
Atmospheric $\nu$ &  $1.9 \times 10^5$ &  $1.2 \times 10^5$ &  $8.5 \times 10^4$ \\
Cosmic $\nu$ &  730 &  420 &  220 \\
\hline
\end{tabular}
\caption{Number of events per year for ARCA for different track selection levels. The selections cover all events passing the trigger conditions (Trigger), events passing the zenith angle $\theta > 80^\circ$ requirement (Zenith) and the final BDT selection (BDT). The expected cosmic event rate is obtained using the flux from equation \ref{eq:cosflux}.}
\label{table:track}
\end{table*}

A stricter cut for horizontal events is applied because of the higher misreconstructed muon contamination. The final selection criteria keep the neutrino purity above 99\% while maximising the signal efficiency. The ARCA event distribution is shown in Figure \ref{fig:trackdistributions} as a function of the reconstructed energy and zenith angle for selected tracks after the cut on the BDT track score. The event rate for all neutrino flavours and interactions for different selection levels is shown in Table \ref{table:track}.

\newpage

\subsection{Shower selection}

Events with a reconstructed shower vertex above the top layer of optical modules are rejected, removing incoming atmospheric muons. This is followed by a cut on the number of hits that fulfill the following hypothesis:
\begin{equation}
    | \Delta t |  = | t_{i} - t_{s} - \frac{d}{c_{\rm water}} | < 20 \text{ ns},
\end{equation}
where $t_i$ is the time of the hit, $t_s$ the reconstructed time of the shower maximum, $d$ the distance between the PMT and the position of the shower maximum and $c_{\rm water}$ the speed of light in water.

To further minimise the atmospheric muon background, a BDT model tailored for showers is trained, focusing on the events that have successfully passed the initial selection requirements. Input variables for training the model are, listed in order of their importance in discriminating signal from background: the estimated track length, the reconstructed position coordinates of both reconstruction procedures, the inertia ratio, the reconstructed zenith angle and the number of hits from both reconstruction procedures. The inertia ratio is a measure of the sphericity of the hit pattern in the detector.

\begin{figure*}
    \centering
    \subfloat[\centering  ]{{\includegraphics[width=0.45\textwidth]{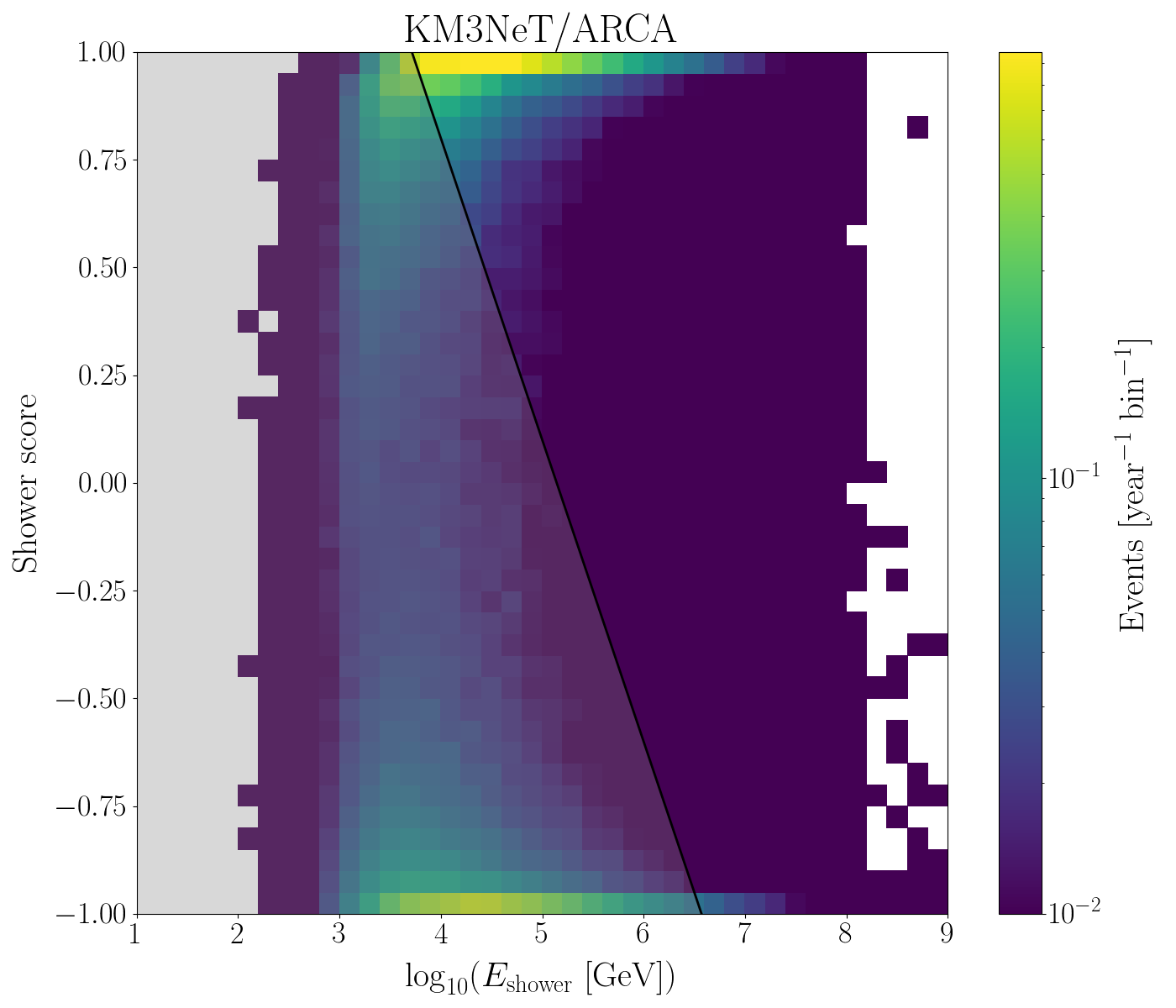}}}%
    \qquad
    \subfloat[\centering  ]{{\includegraphics[width=0.45\textwidth]{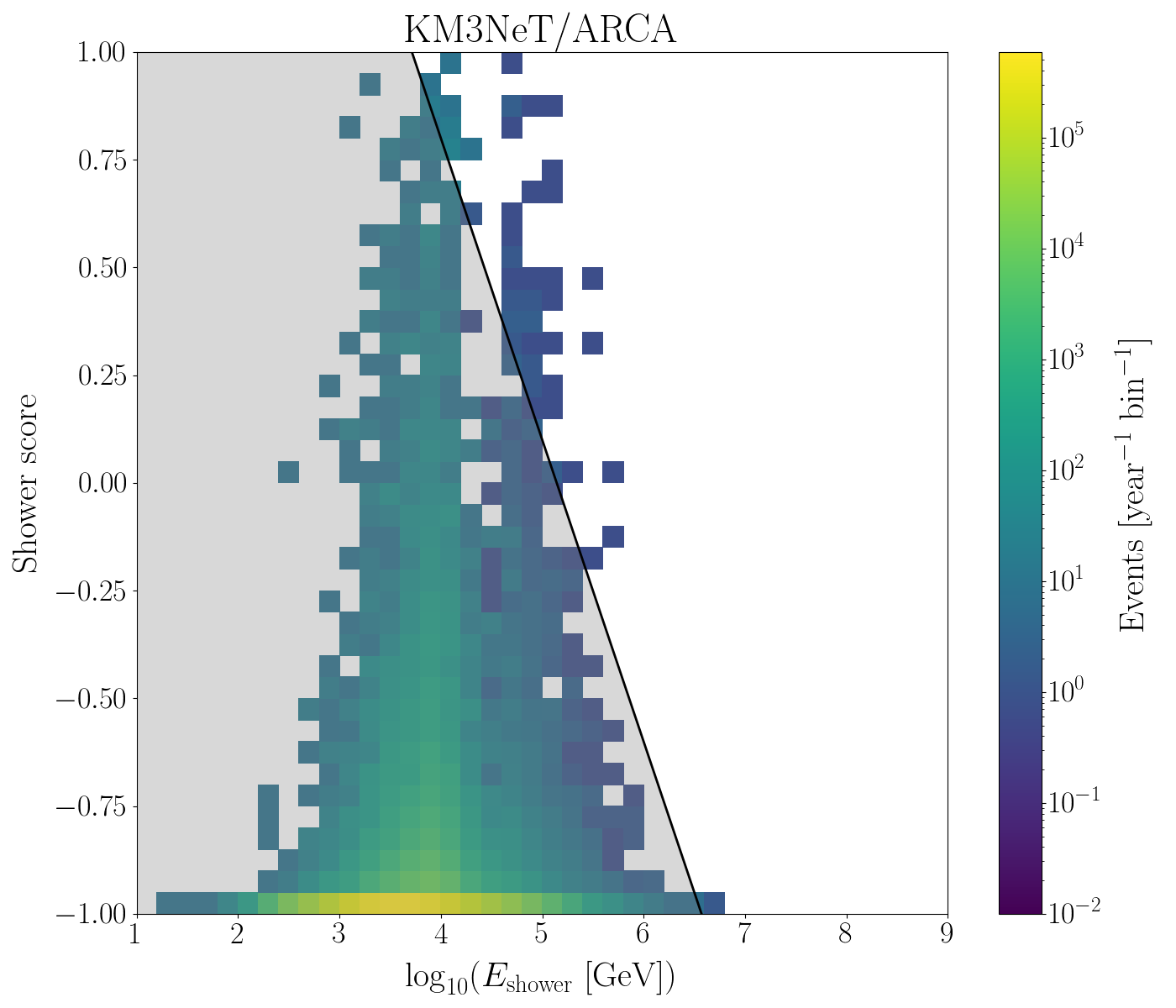} }}%
    \caption{ARCA event distribution as a function of the shower classifier score and the reconstructed energy for cosmic $\nu_e$ + $\bar{\nu}_e$ CC events (a) and the atmospheric muon background (b) after the requirement on the number of hits. The shaded region covers events that are rejected. The expected cosmic event rate is obtained using the flux from equation \ref{eq:cosflux}.}
    \label{fig:showerbdt}
\end{figure*}

The final selection is based on a two-dimensional cut using the reconstructed shower energy and the \textit{shower score} from the BDT. The rate of cosmic neutrinos from $\nu_e$ CC interactions and the atmospheric muon background are shown in Figure \ref{fig:showerbdt} as a function of the reconstructed shower energy and shower score. The shaded region covers events that are rejected.

\begin{figure*}
    \centering
    \subfloat[\centering  ]{{\includegraphics[width=0.45\textwidth]{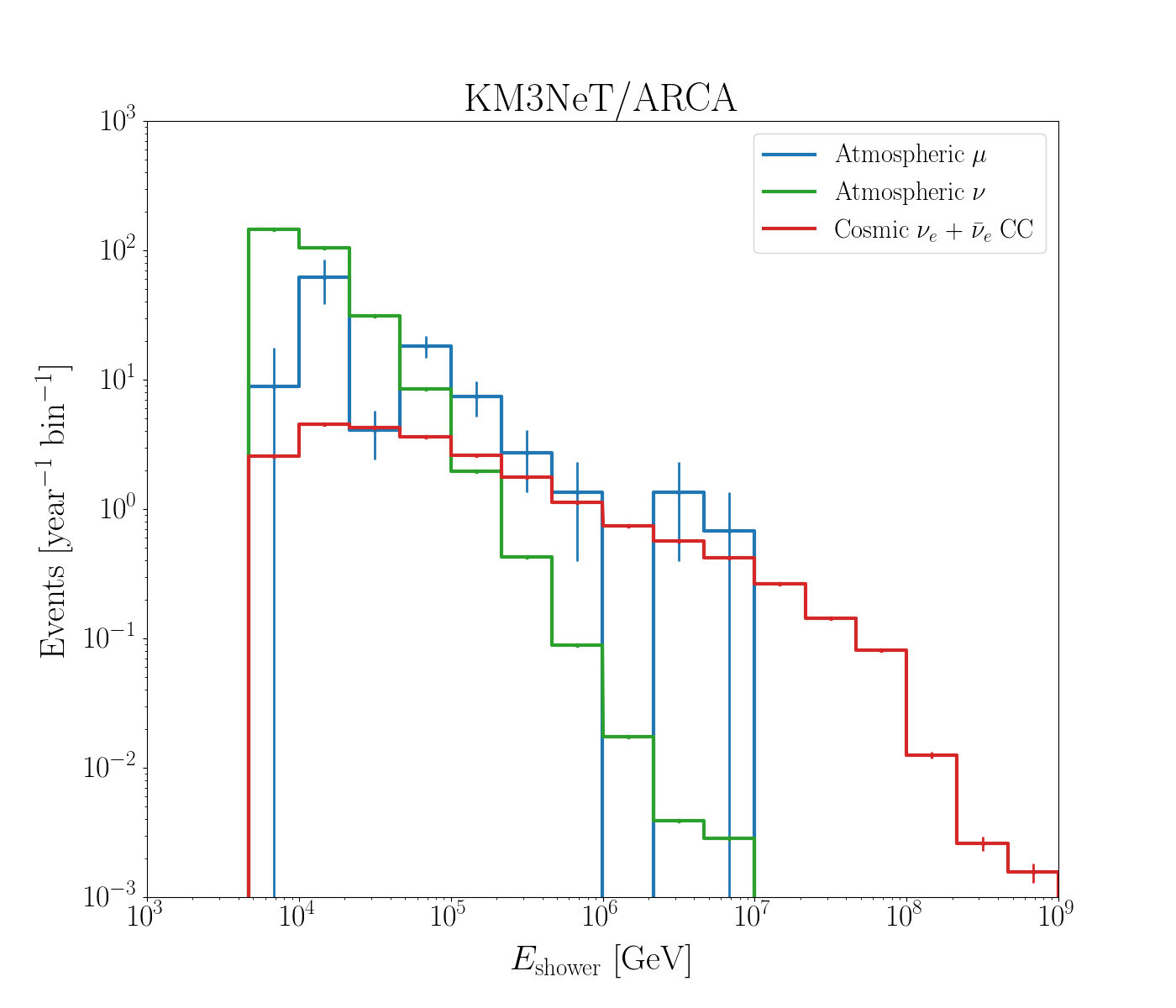}}}%
    \qquad
    \subfloat[\centering  ]{{\includegraphics[width=0.45\textwidth]{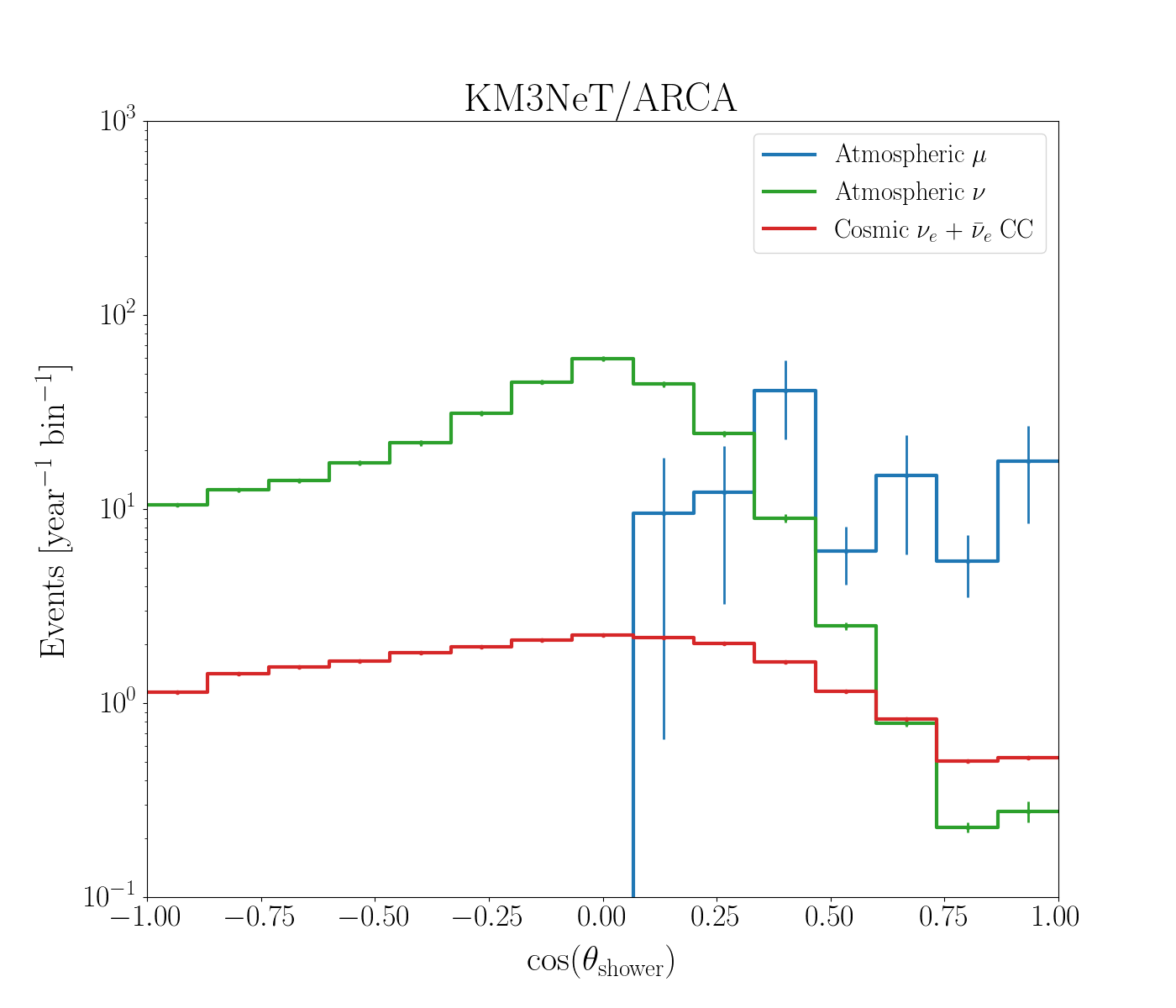} }}%
    \caption{Event rates per year versus the reconstructed energy (a) and zenith angle (b) for the shower selection. The expected cosmic event rate is obtained using the flux from equation \ref{eq:cosflux}.}
    \label{fig:showerdistributions}%
\end{figure*}

The shower-like event rate is shown in Figure \ref{fig:showerdistributions} as a function of the reconstructed energy and zenith angle of the selected events. The event rates for all neutrino flavours and interactions for different selection levels are presented in Table \ref{table:shower}.

\begin{table*}
\centering
\begin{tabular}{lllll}
\hline
  &  Trigger [yr$^{-1}$] &  Containment [yr$^{-1}$] &  Hits [yr$^{-1}$] &  BDT [yr$^{-1}$] \\
\hline
Atmospheric $\mu$ (> 10 TeV) &  $8.1 \times 10^7$ &  $1.9 \times 10^7$ &  $3.2 \times 10^6$ &  110\\
Atmospheric $\nu$ &  $1.9 \times 10^5$ &  $8.0 \times 10^4$ &  $1.2 \times 10^4$ &  290\\
Cosmic $\nu$ &  730 &  370 &  200 &  60\\
\hline
\end{tabular}
\caption{Number of events per year for ARCA for different shower selection levels. The columns represent all events passing the trigger conditions (Trigger), the events that pass the containment criteria (Containment), the events that pass the hit requirements (Hits) and the final BDT selection (BDT). The expected cosmic event rate is obtained using the flux from equation \ref{eq:cosflux}.}
\label{table:shower}
\end{table*}

\subsection{Performances}

\begin{figure*}
    \centering
    \subfloat[\centering  ]{{\includegraphics[width=0.55\textwidth]{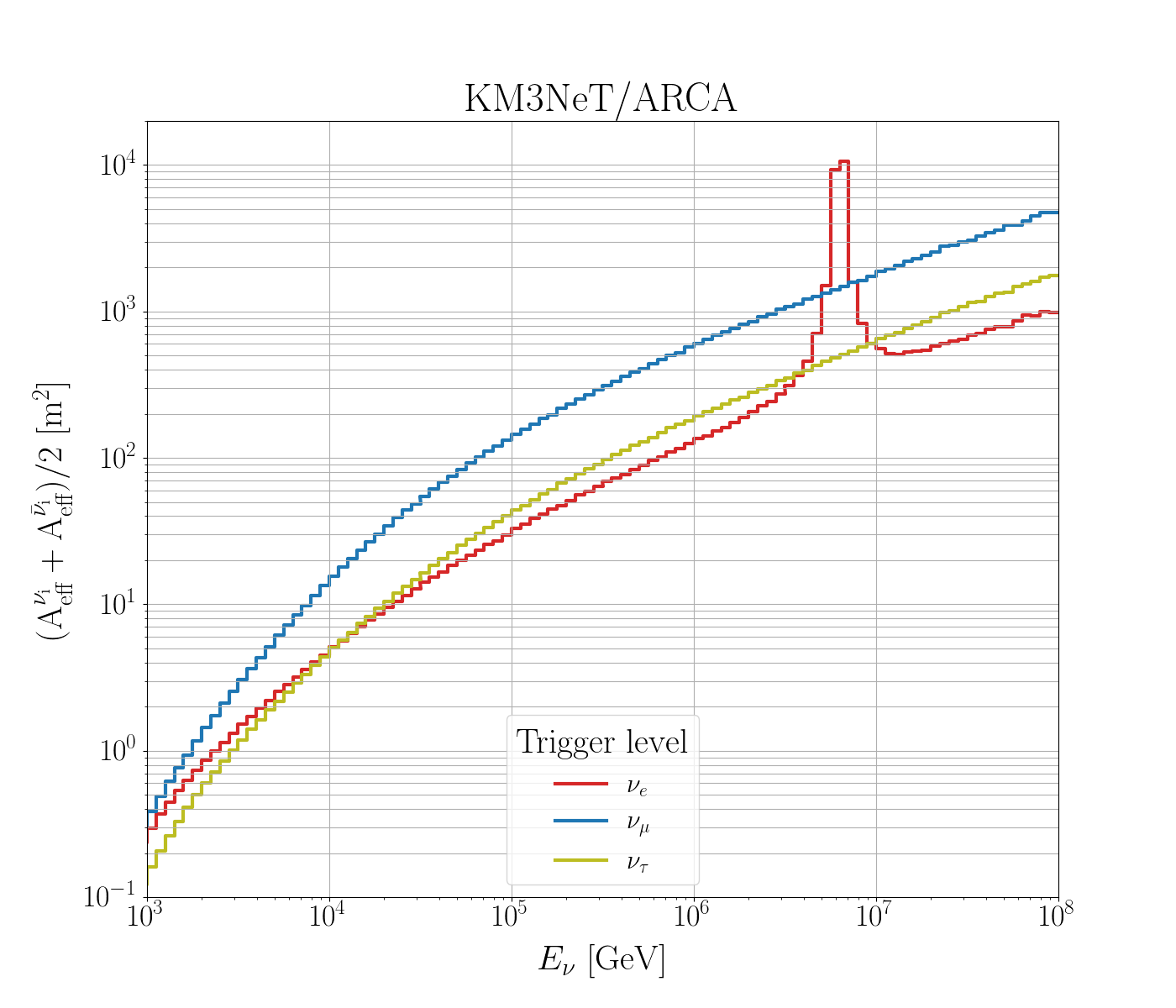}}}%
    \qquad
    \subfloat[\centering  ]{{\includegraphics[width=0.45\textwidth]{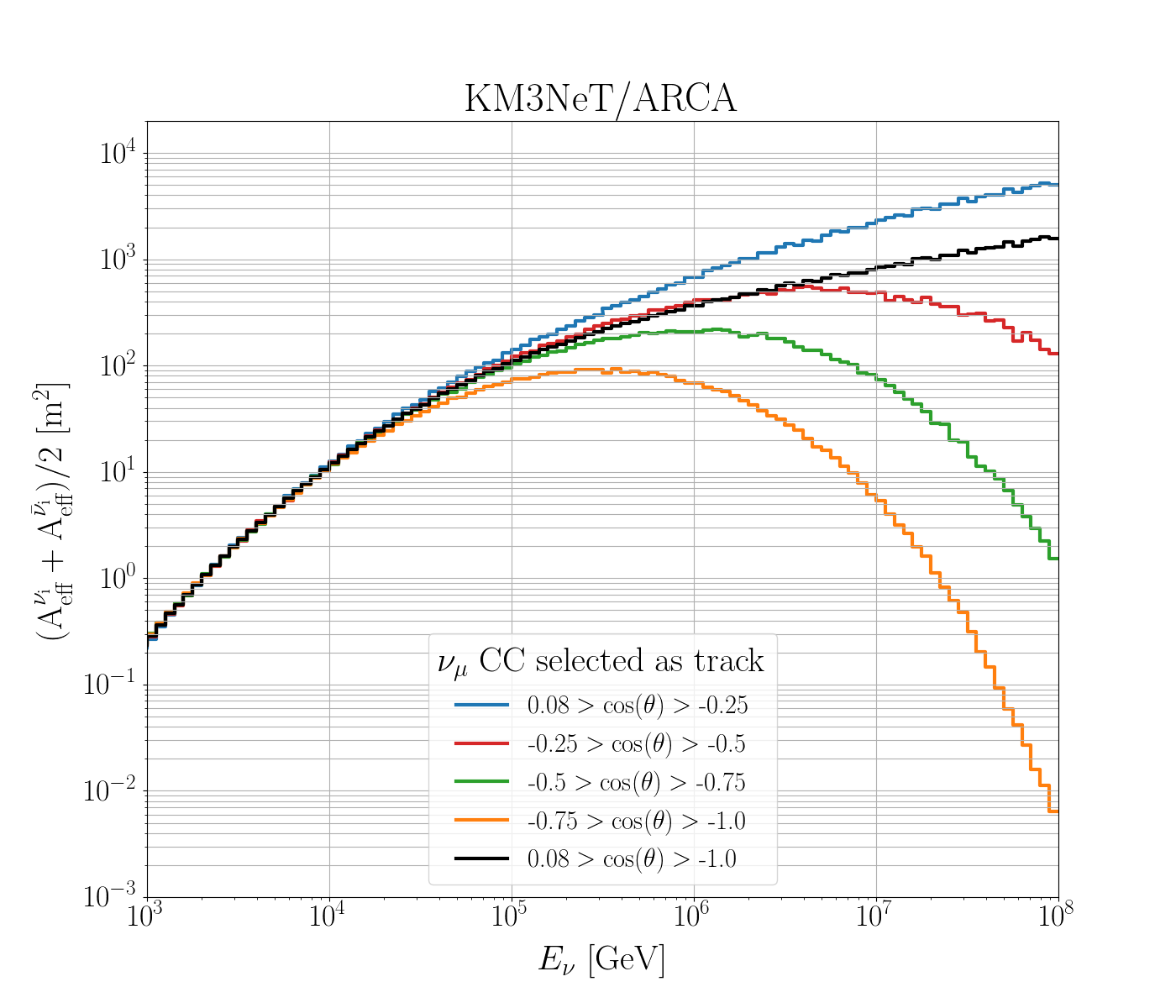} }}%
    \subfloat[\centering  ]{{\includegraphics[width=0.45\textwidth]{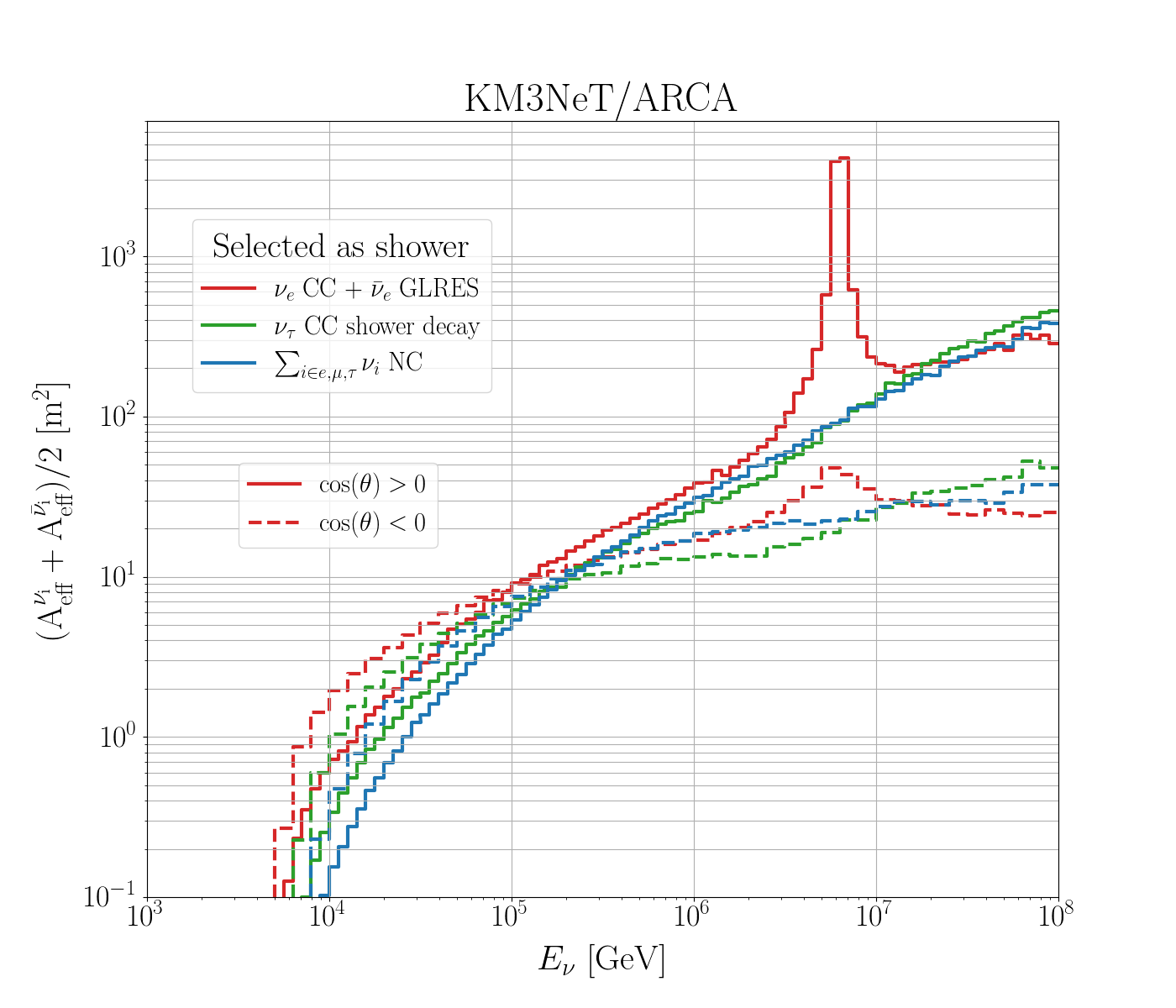} }}%
    \caption{The ARCA effective area for a flux of $\nu_i + \bar{\nu}_i$ at trigger level for all neutrino flavours and interactions (a). The effective area for $\nu_\mu$ CC events selected as track for different $\cos(\theta)$ ranges (b). The effective area for the shower channel covers both upgoing and downgoing events from $\nu_e$ CC, $\nu_\tau$ CC and $\nu$ NC interactions (c). }
    \label{fig:aeff}%
\end{figure*}

The effective area $A_{\rm eff}^{(\nu_i+\bar{\nu}_i)/2}$ at trigger level, and for the final track and shower selections are shown in Figure \ref{fig:aeff} for different neutrino flavours and $\cos(\theta)$ ranges. This quantity is defined as the ratio between the detected event rate $N_{\rm det}$ and the neutrino flux $\Phi$. It represents the area of a hypothetical detector that has a 100\% neutrino detection efficiency. The effective area is averaged over particles and anti-particles such that each flavour can be convolved with a $\Phi^{\nu_i+\bar{\nu}_i}$ flux in order to obtain the event rate. The decrease of effective area for upgoing events is due to the absorption of high-energy neutrinos in the Earth. The distribution for the shower channel shows an increase of effective area at a neutrino energy of 6.3 PeV due the resonant production of a $W^-$ boson when an $\bar{\nu}_e$ interacts with an electron \cite{glashow1960resonant}. 

\begin{figure*}
    \centering
    \subfloat[\centering  ]{{\includegraphics[width=0.45\textwidth]{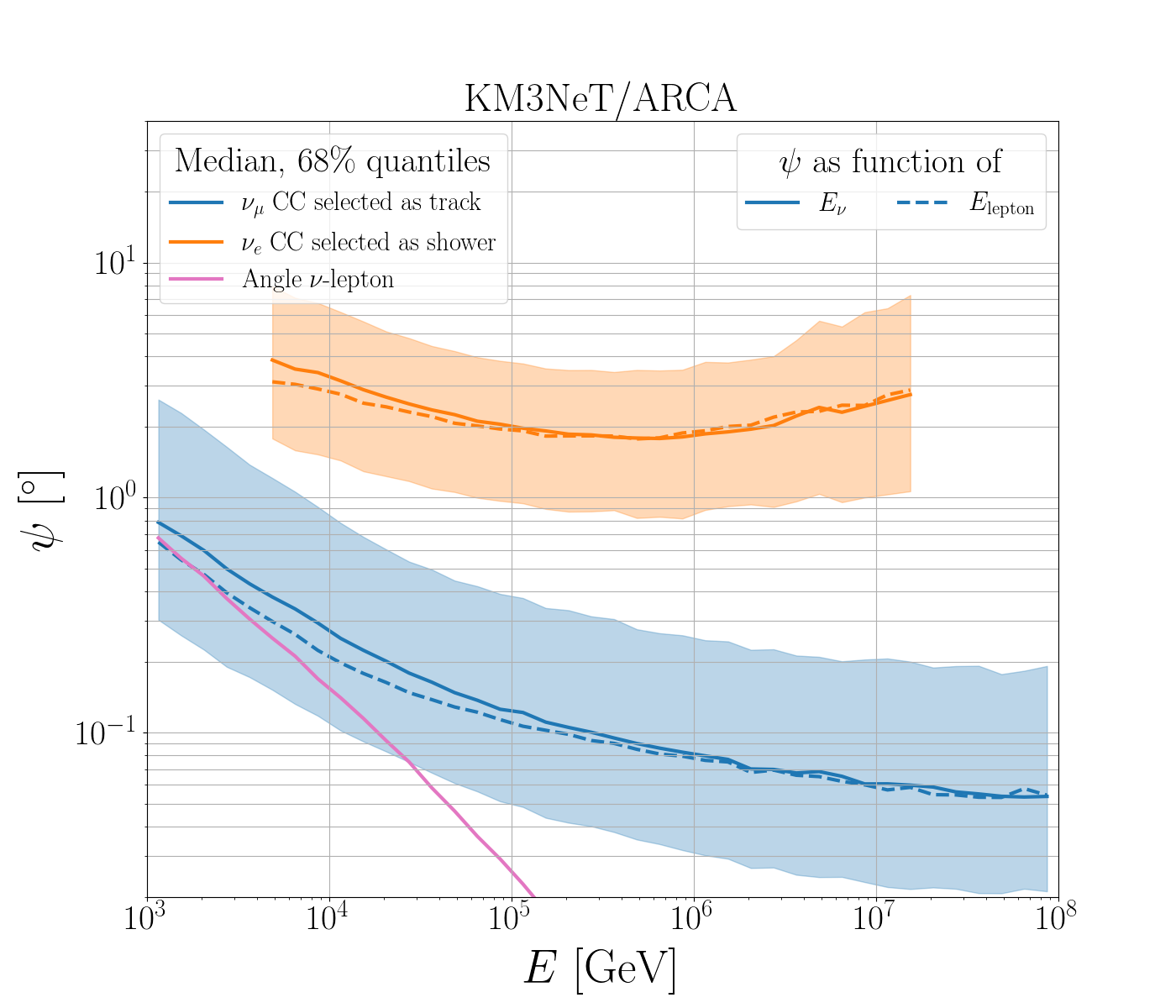}}}%
    \qquad
    \subfloat[\centering  ]{{\includegraphics[width=0.45\textwidth]{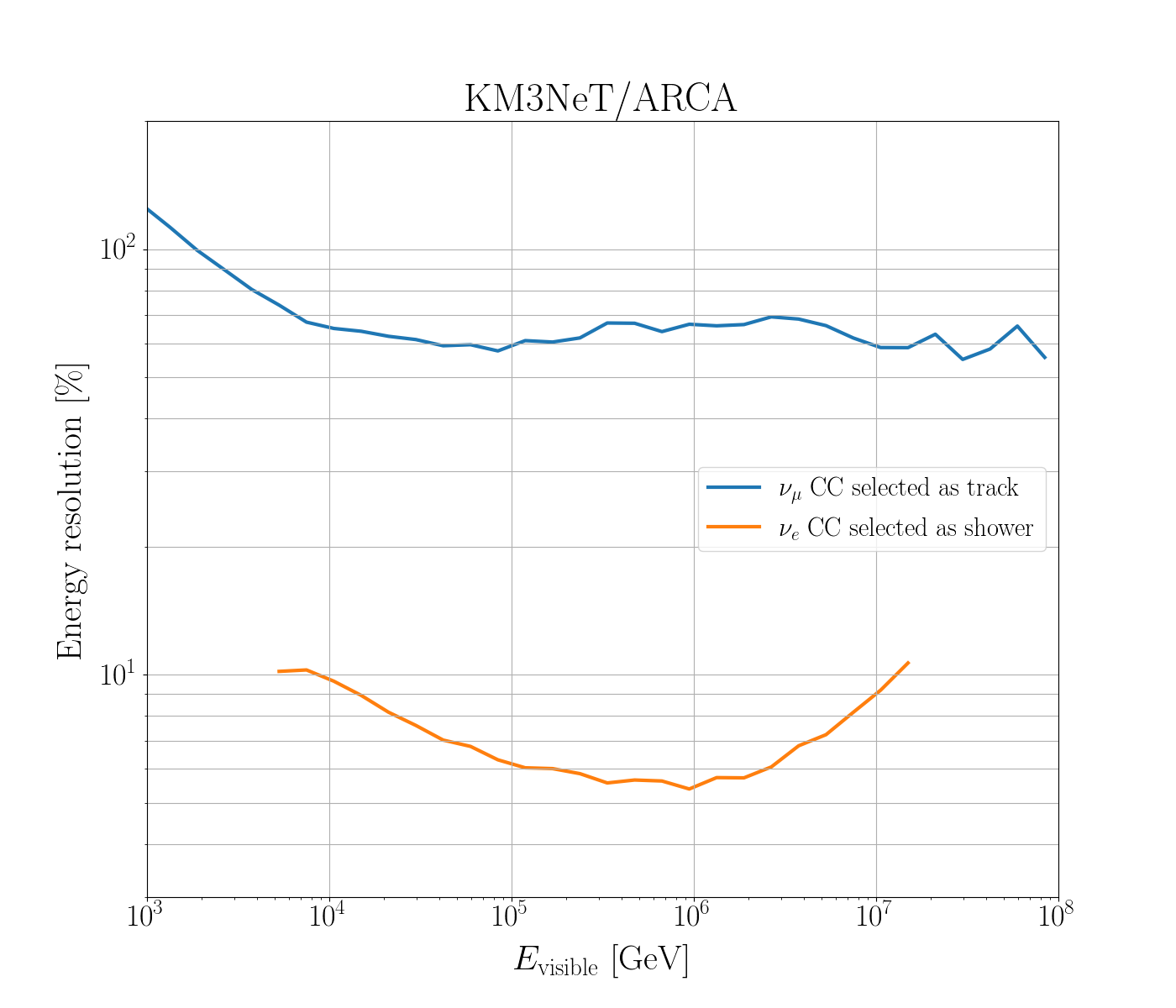} }}%
    \caption{ARCA angular deviation (a) and energy resolution (b) for $\nu_\mu$ CC events selected as track and for $\nu_e$ CC events selected as shower. The angular deviation is shown alongside the median angle between the neutrino and the outgoing lepton in CC interactions. }
    \label{fig:resolution}%
\end{figure*}

In Figure \ref{fig:resolution}a, the angular deviation $\psi$ is depicted. This quantity represents the difference between the reconstructed direction and the true neutrino direction. Specifically, this quantity is illustrated for $\nu_\mu$ CC events identified as tracks and for $\nu_e$ CC events identified as showers. The median resolution reaches 0.1$^\circ$ for tracks and below 2$^\circ$ for showers at 300 TeV. The energy resolution is shown in Figure \ref{fig:resolution}b where $E_{\rm visible}$ is the sum of energies of all particles producing light in the detector. Saturation of the PMTs is responsible for the deterioration in the performance of direction and energy reconstruction in the shower reconstruction. The final event rates and neutrino purity for the full detector are shown in Table \ref{table:rate}.
    
\begin{table*}
\centering
\begin{tabular}{llll}
\hline
                  & Trigger [yr$^{-1}$] & Track selection [yr$^{-1}$] & Shower selection [yr$^{-1}$] \\ 
\hline
Atmospheric $\mu$ (> 10 TeV) &  $8.1 \times 10^7$  & 714             & 110              \\ 
Atmospheric $\nu$ & $1.9 \times 10^5$  & $8.5 \times 10^4$          & 290             \\ 
Cosmic $\nu$      & 730     & 220             & 60              \\ 
\hline
Neutrino purity & & 99\% & 77\% \\ 
\hline
\end{tabular}
\caption{Event rate per year for ARCA at trigger level and after applying the requirements for the track and shower selections. The expected cosmic event rate is obtained using the flux from equation \ref{eq:cosflux}.}
\label{table:rate}
\end{table*}

%% file: point_sources.tex
\section{Point-like source search}

Simulated events classified as either tracks or showers from the selection procedure described above are used to create a set of detector response functions that include the effective area, the point spread function and the energy response. Should an event be classified both as a track and as a shower, it is allocated to the track channel because of its superior directional reconstruction performance. The energy response translates the true neutrino energy distribution into the reconstructed energy distribution, while the point spread function describes how neutrino directions are smeared by the reconstruction. These functions are used as inputs to the estimation of the sensitivity and discovery potential of ARCA in the search for point-like neutrino sources.

\subsection{Point spread function}

\begin{figure*}
    \centering
    \subfloat[\centering  ]{{\includegraphics[width=0.45\textwidth]{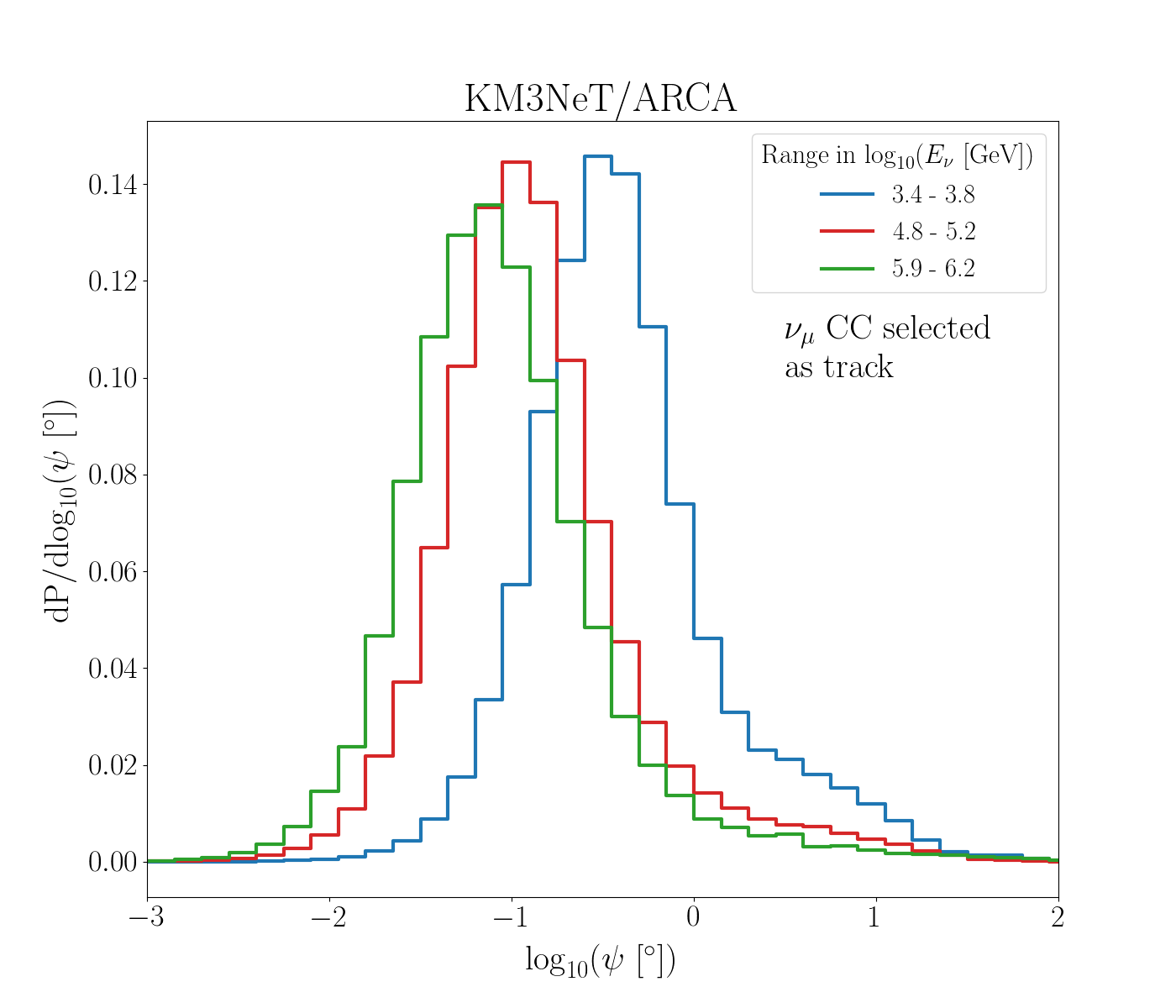}}}%
    \subfloat[\centering  ]{{\includegraphics[width=0.45\textwidth]{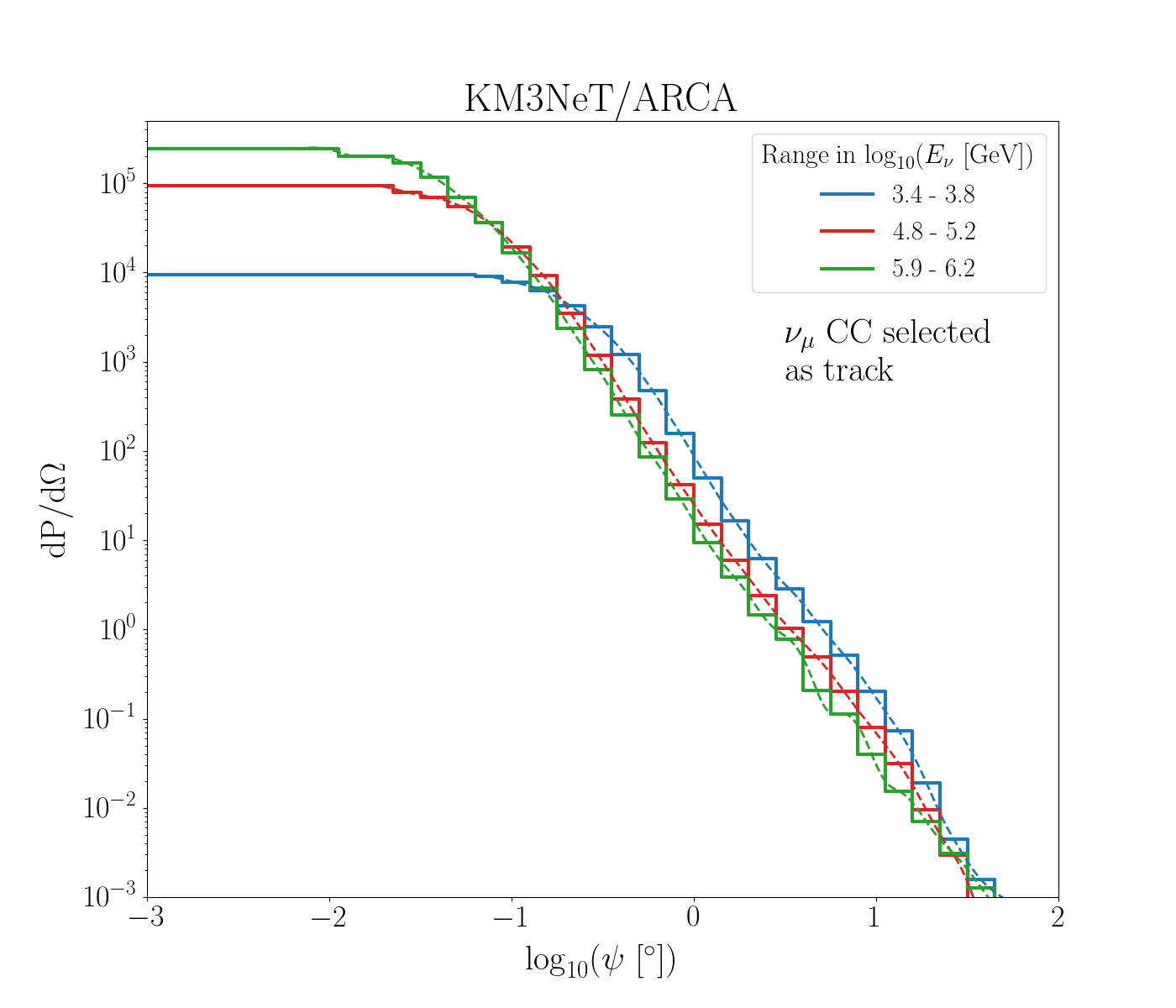} }}%

    \qquad

    \subfloat[\centering  ]{{\includegraphics[width=0.45\textwidth]{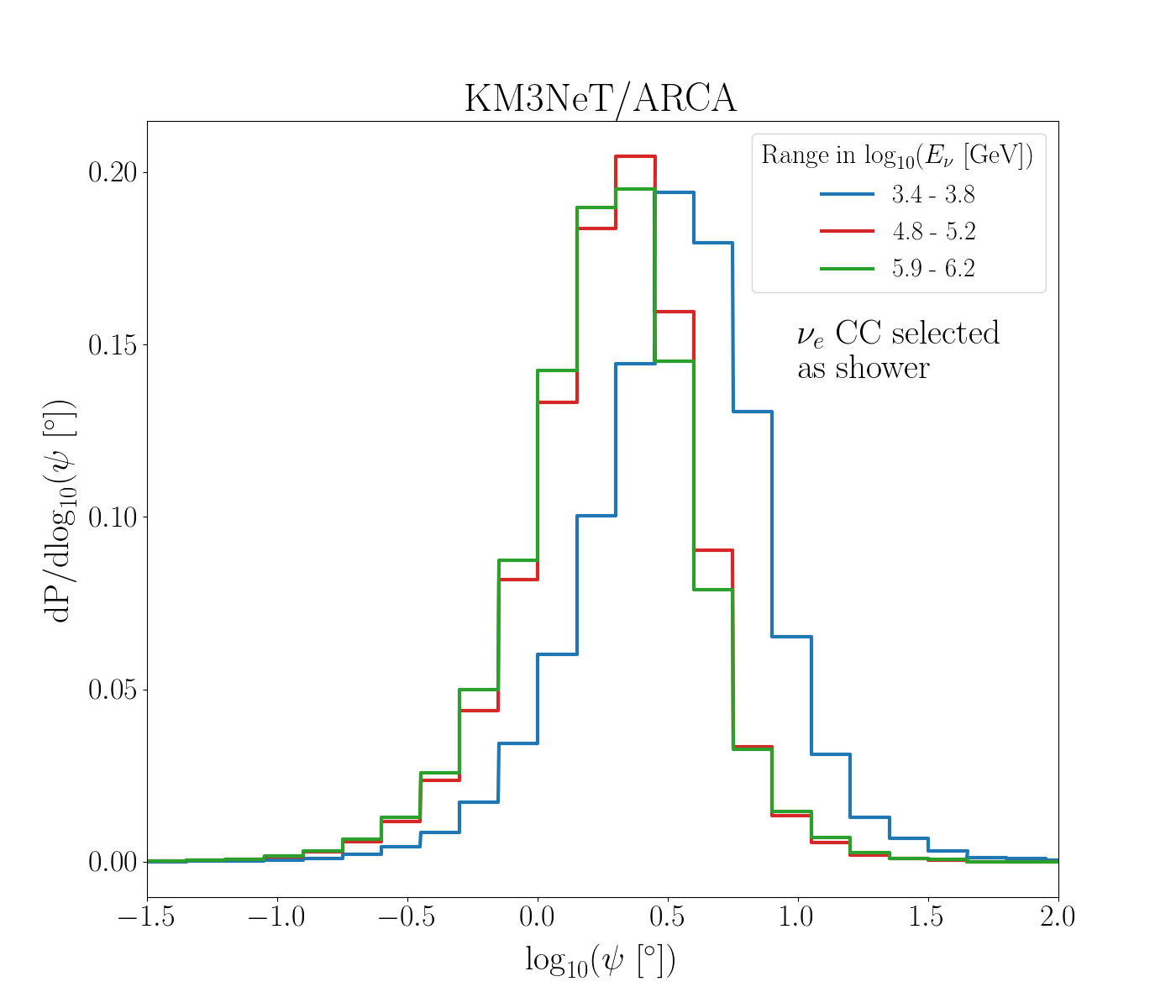}}}%
    \subfloat[\centering  ]{{\includegraphics[width=0.45\textwidth]{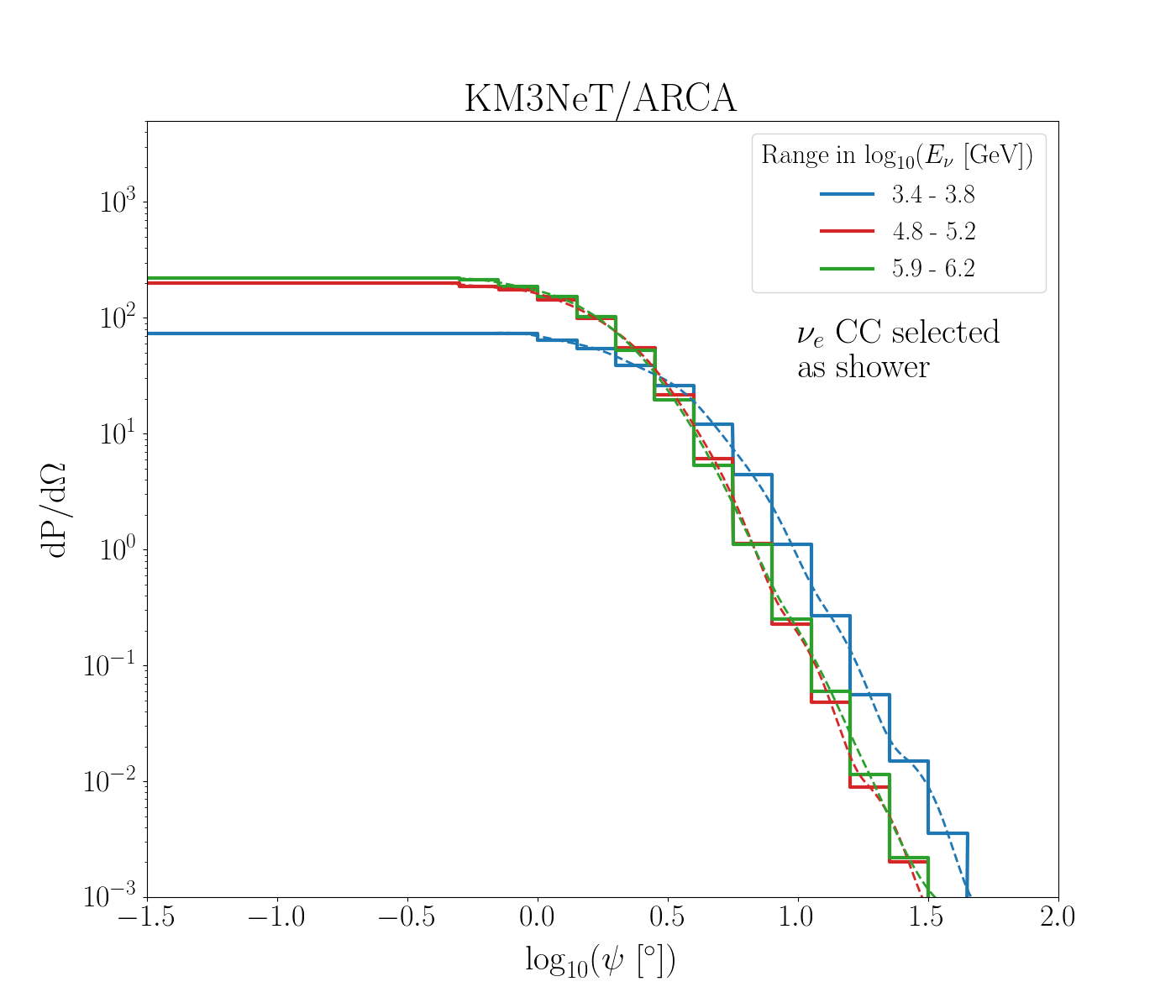} }}%
    \caption{Distributions of $dP/d\log_{10}(\psi)$ (a, c) and $dP/d\Omega$ (b, d) for selected energy ranges. The distributions are shown for $\nu_\mu$ CC events selected as track (a, b) and for $\nu_e$ CC events selected as shower (c, d).}
    \label{fig:psf}%
\end{figure*}

The angular deviation $\psi$ between the reconstructed and true direction of the selected neutrino events is used to construct the point spread function. The point spread function is defined as the event density per unit solid angle ($dP/d\Omega$) as a function of $\psi$. Distributions of $dP/d\log_{10}(\psi)$ are obtained from the Monte Carlo simulations and converted to $dP/d\Omega$ via
\begin{align}
    & \frac{dP}{d\Omega} = \frac{d\log_{10}(\psi)}{d\Omega} \frac{dP}{d\log_{10}(\psi)} = \\
    & \frac{1}{2 \psi \pi \sin (\psi) \log(10)} \frac{dP}{d\log_{10}(\psi)} \notag
\end{align}
for each neutrino flavour and observation channel. This function is constructed for different neutrino energy ranges accounting for the energy dependence of the reconstruction. Example distributions of $dP/d\log_{10}(\psi)$ and $dP/d\Omega$ for $\nu_\mu$ CC events selected as track and $\nu_e$ CC events selected as shower are shown in Figure \ref{fig:psf}.

\subsection{Method}

The sensitivity and discovery potential estimations are obtained using a \textit{binned} method where the datasets are represented by event rate distributions as a function of reconstructed energy and angular distance to the source $\psi$. The detector response functions are used to create expected distributions for signal and background for different flux models and periods of data collection. The signal flux models are characterised by a single power law with different spectral indices $\gamma$ according to
\begin{equation}
    \Phi^{\nu_i+\bar{\nu}_i} = \Phi_0 \Big( \frac{E}{\text{GeV}} \Big)^{-\gamma},
\label{eq:flux_point}
\end{equation}
where $i=e,\mu,\tau$ and spectral index $\gamma=2.0,2.5,3.2$. The spectral index ranges between 2.0 which matches the energy spectrum of cosmic rays undergoing Fermi acceleration \cite{fermi1949origin} to 3.2 which was is the best-fit determined by the IceCube Collaboration from the observation of the nearby active galaxy NGC 1068 \cite{icecube2022evidence}.

The flux models are convolved with the effective area, the point spread function and the energy response to obtain two-dimensional distributions for the expected signal ($S_ij$) for each channel. The background distributions ($B_ij$) are obtained by calculating the density of background events per declination ($\delta$) band from atmospheric neutrinos and muons. The flux normalisation $\Phi_0$ is scaled with a varying signal strength $\zeta$ to study the sensitivity and discovery potential of ARCA.

\begin{figure*}
    \centering
    \subfloat[\centering  ]{{\includegraphics[width=0.45\textwidth]{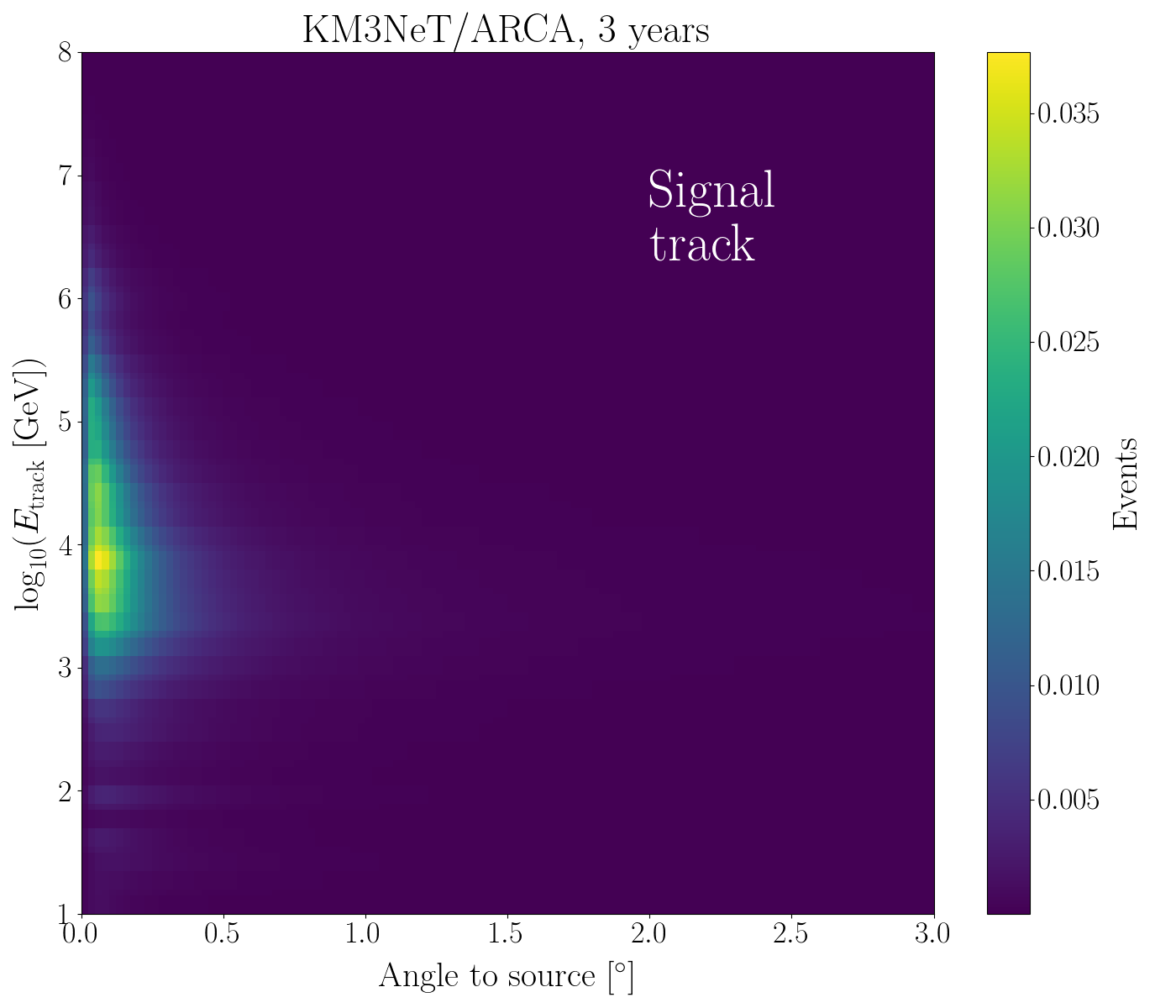}}}%
    \subfloat[\centering  ]{{\includegraphics[width=0.45\textwidth]{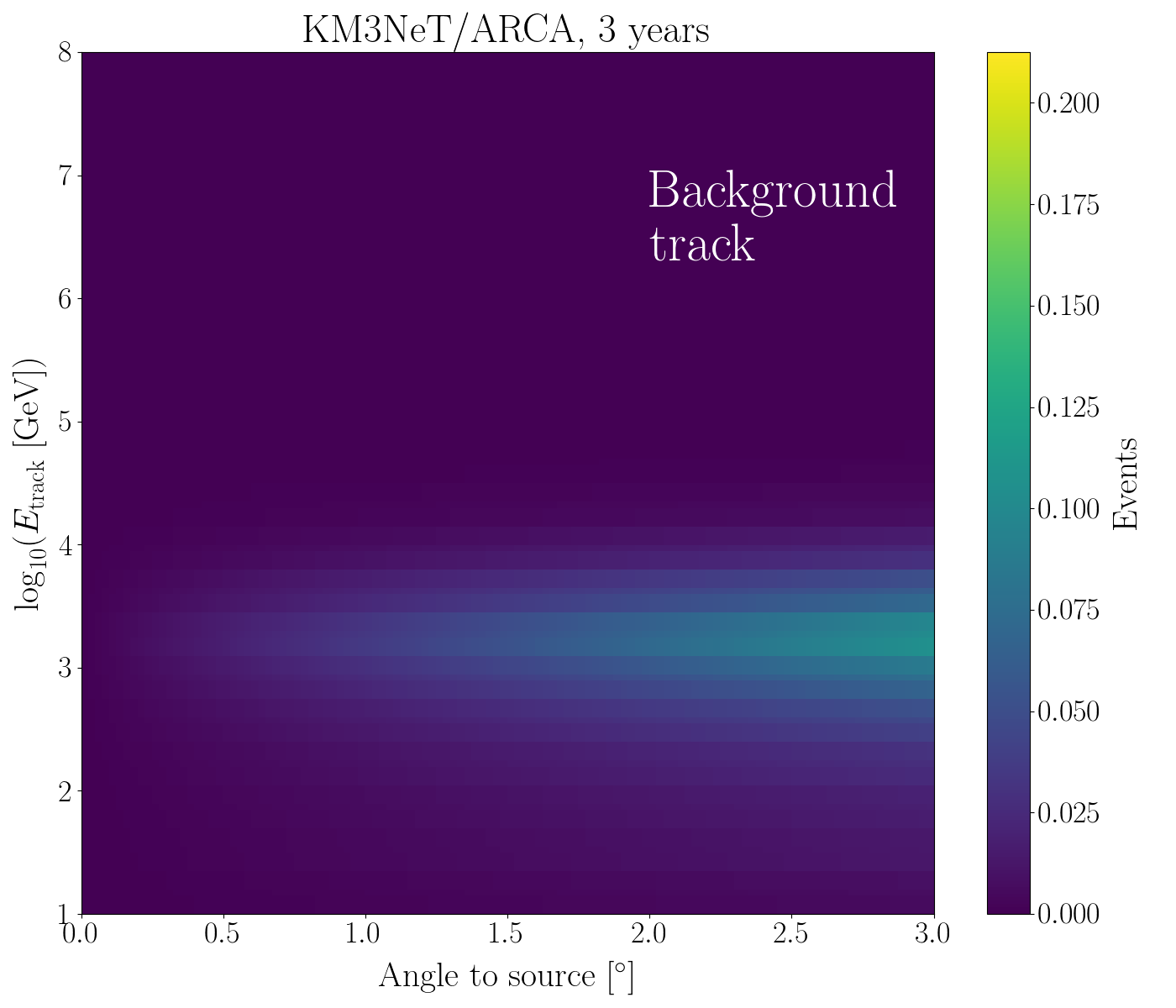} }}%

    \qquad

    \subfloat[\centering  ]{{\includegraphics[width=0.45\textwidth]{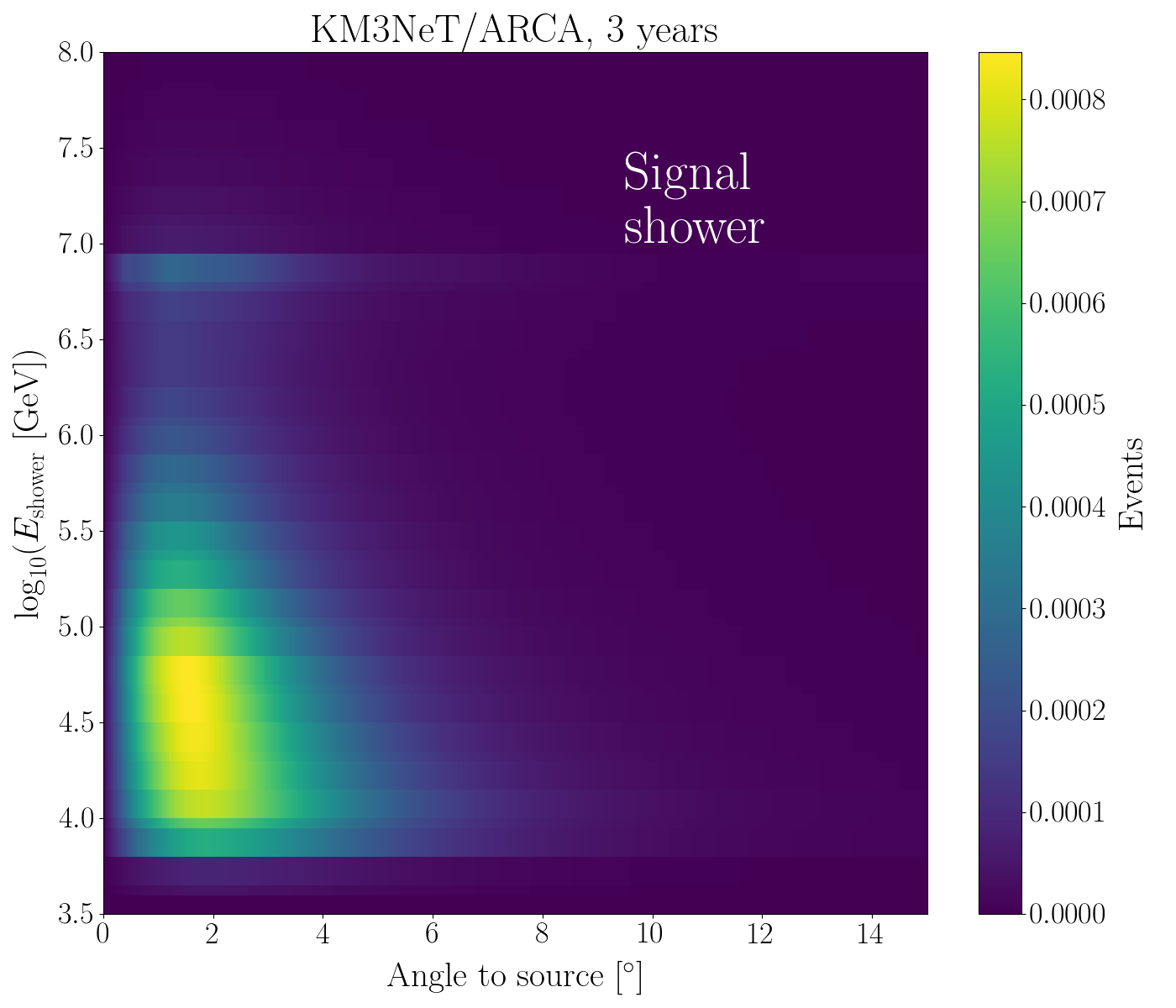}}}%
    \subfloat[\centering  ]{{\includegraphics[width=0.45\textwidth]{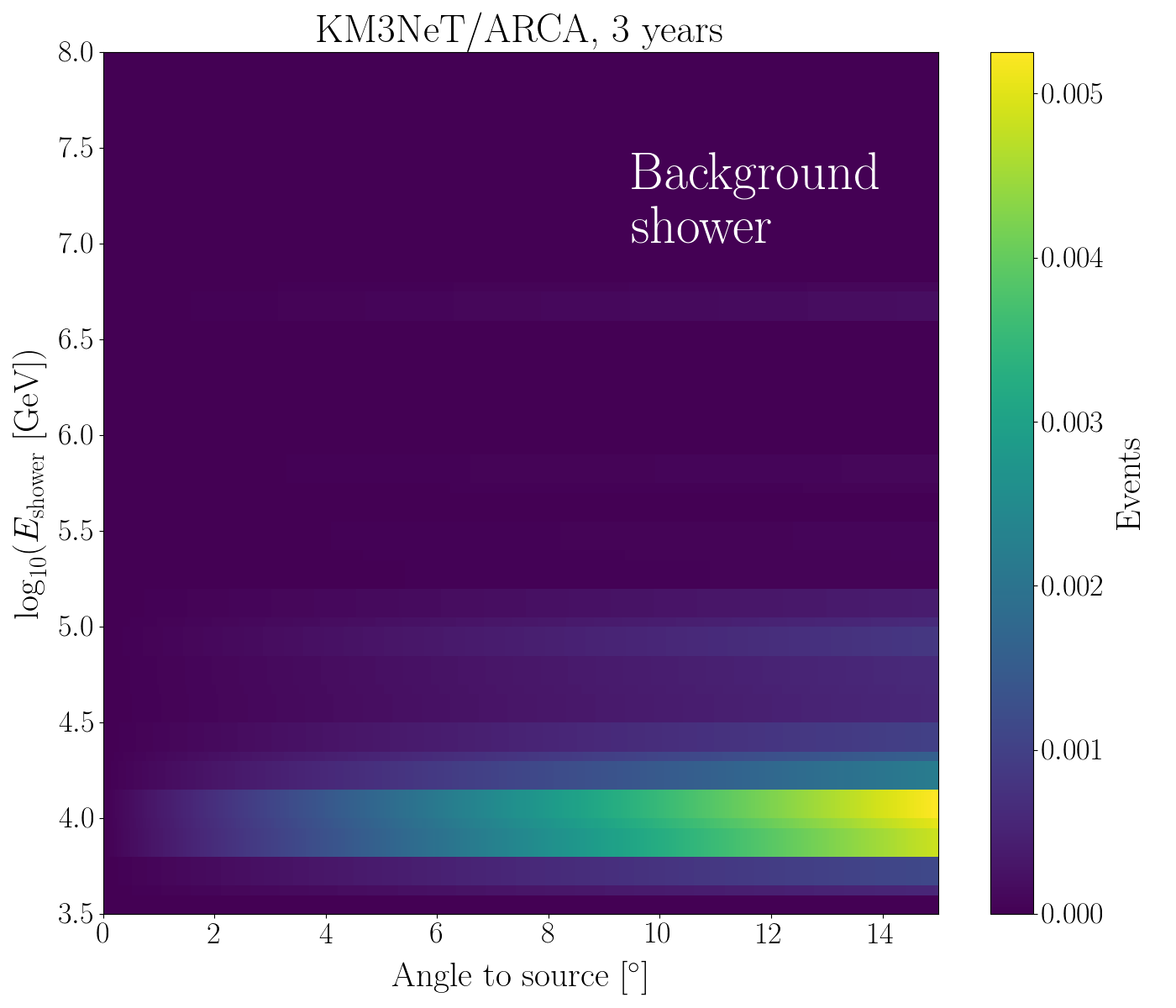} }}%
    \caption{Expected signal and background histograms for three years of ARCA operation looking at a source at sin($\delta$) = 0.1 using $\gamma = 2.0$ and $\Phi_0 = 4 \times 10^{-9} \text{ GeV}^{-1} {\rm s}^{-1} {\rm cm}^{-2}$. The event rate distributions are represented as a function of reconstructed energy and distance to the source $\psi$. The left column shows the signal histograms and the right column the background histograms. The upper row contains the track channel and the bottom row the shower channel. The covered solid angle of the sky increases linearly with the angle to the source.}
    \label{fig:sig_bg_point}%
\end{figure*}

An example of expected signal and background distributions for tracks and showers is shown in Figure \ref{fig:sig_bg_point}, for a source at sin($\delta$) = 0.1, $\gamma =2.0$, $\Phi_0 = 4 \times 10^{-9} \text{ GeV}^{-1} {\rm s}^{-1} {\rm cm}^{-2}$ and three years of ARCA operation. 

Based on the expected signal and background distributions, pseudo-experiments using Poisson statistics are generated. One pseudo-dataset contains two histograms of observed events: one for the track channel and one for the shower channel. The expectation value for the number of events of a pseudo-dataset $N_{ij}^c$ for a given channel $c$ and bin $ij$ is defined as
\begin{equation}
    N_{ij}^c = B_{ij}^c + \zeta S_{ij}^c.
\end{equation}
The signal strength $\zeta$ is varied from 0 for background-only (H0) datasets to higher values for combined signal and background (H1) datasets. The potential to discriminate H1 from H0 datasets is studied by choosing a test statistic $\lambda$. The test statistic is a log likelihood ratio defined as
\begin{equation}
\label{eq:ts}
    \lambda = \log \mathcal{L} (\zeta = \hat{\zeta}) - \log \mathcal{L} (\zeta = 0),
\end{equation}
where the log likelihood for a H0 hypothesis is subtracted from the log likelihood for a H1 hypothesis. The estimated signal strength $\hat{\zeta}$ of the dataset is obtained by maximising the likelihood
\begin{align}
    & \log \mathcal{L} = \\
    & \sum_{ij \in \text{bins}} \sum_{c \in \text{channels}} N_{ij}^c \log(B_{ij}^c + \zeta S_{ij}^c) - B_{ij}^c - \zeta S_{ij}^c, \notag
\end{align}
while varying $\zeta$.

\begin{figure*}
\centering
\includegraphics[width=0.45\textwidth]{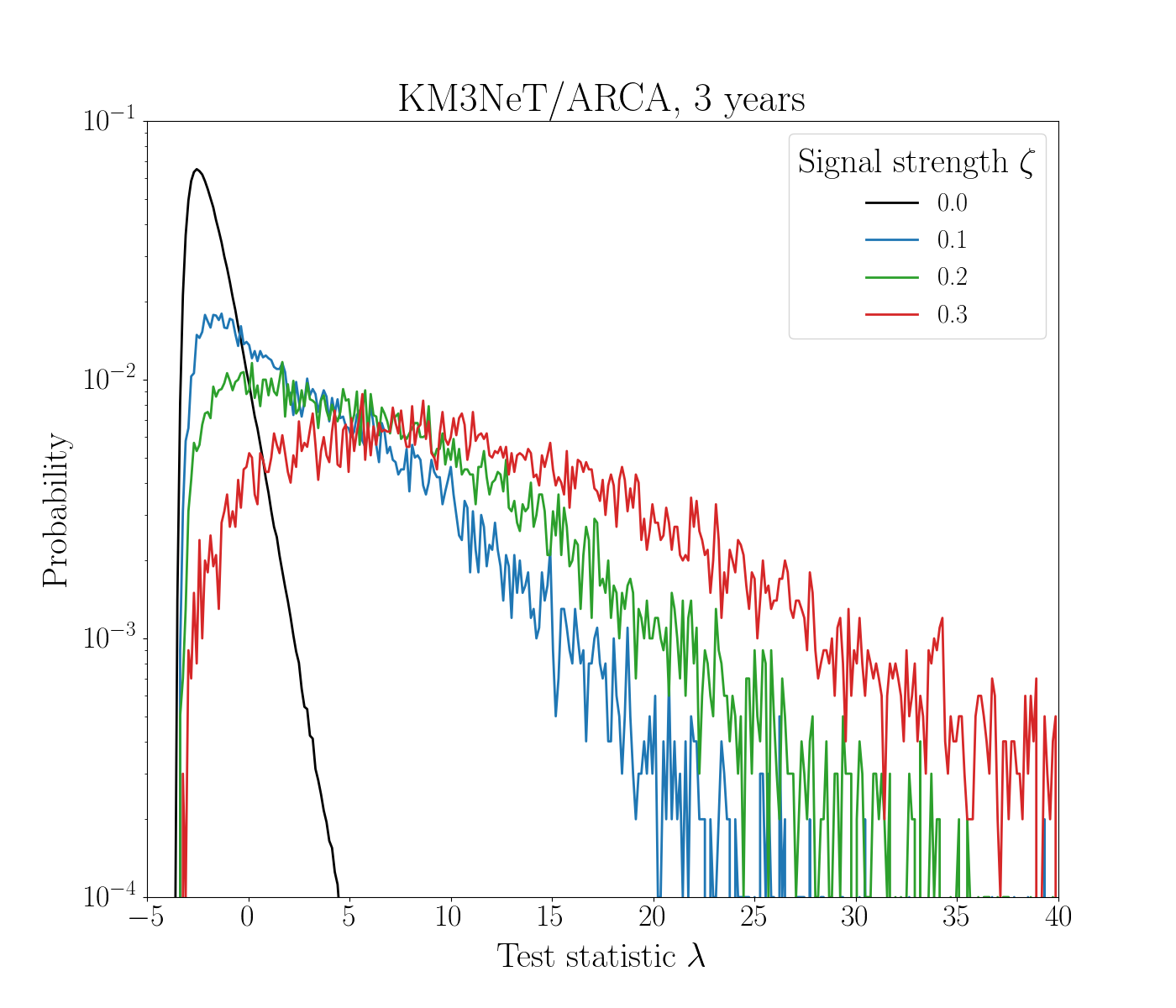}
\caption{Test statistic distribution for a point-like source at sin($\delta$) = 0.1 using $\Phi_0 = 4 \times 10^{-9} \text{ GeV}^{-1} {\rm s}^{-1} {\rm cm}^{-2}$ and $\gamma = 2.0$ for three years of ARCA operation.}
\label{fig:tstrack}
\end{figure*}

The test statistic for different signal strengths when looking at the example source introduced earlier in this section using tracks and showers is shown in Figure \ref{fig:tstrack}.

\begin{figure*}
    \centering
    \subfloat[\centering  ]{{\includegraphics[width=0.45\textwidth]{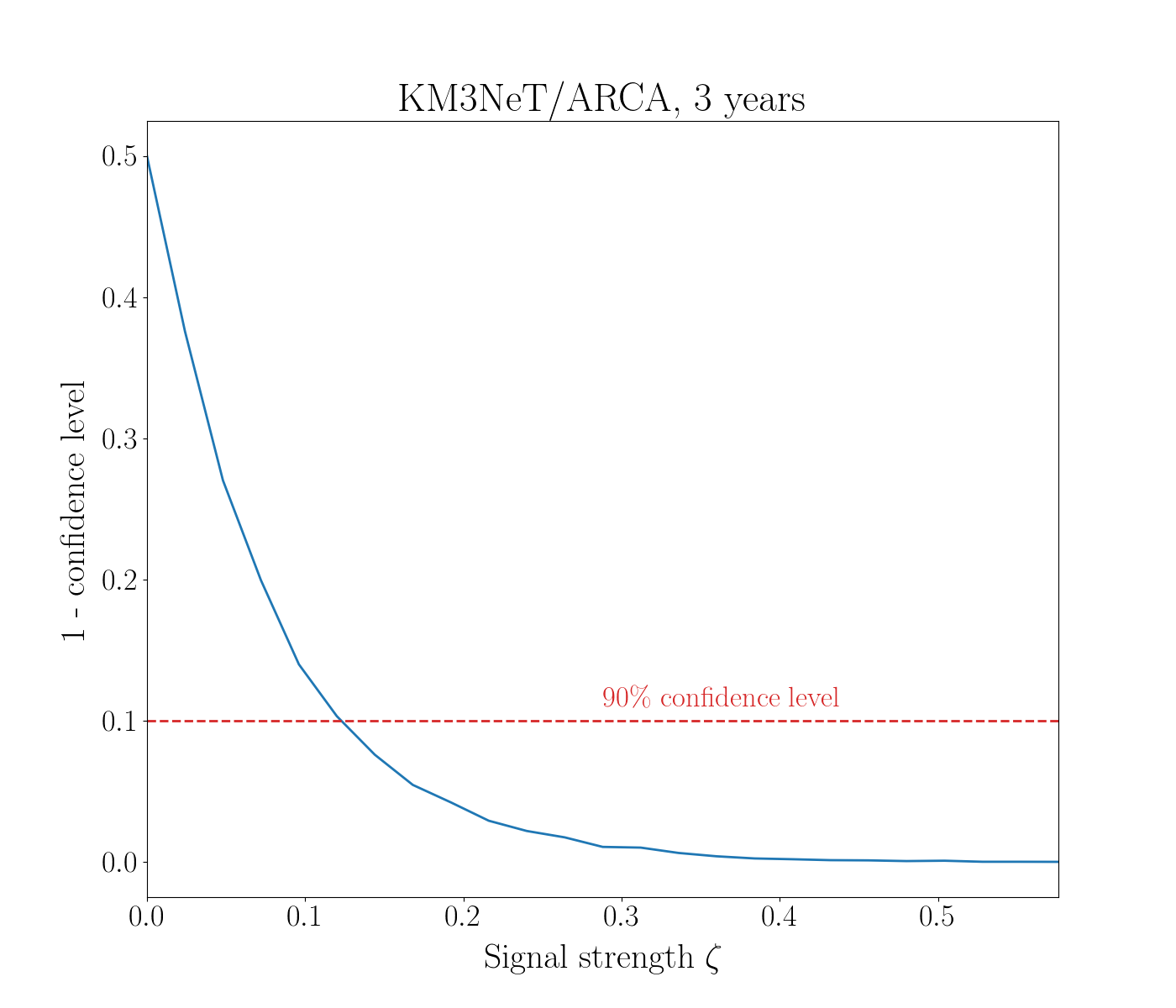}}}%
    \qquad
    \subfloat[\centering  ]{{\includegraphics[width=0.45\textwidth]{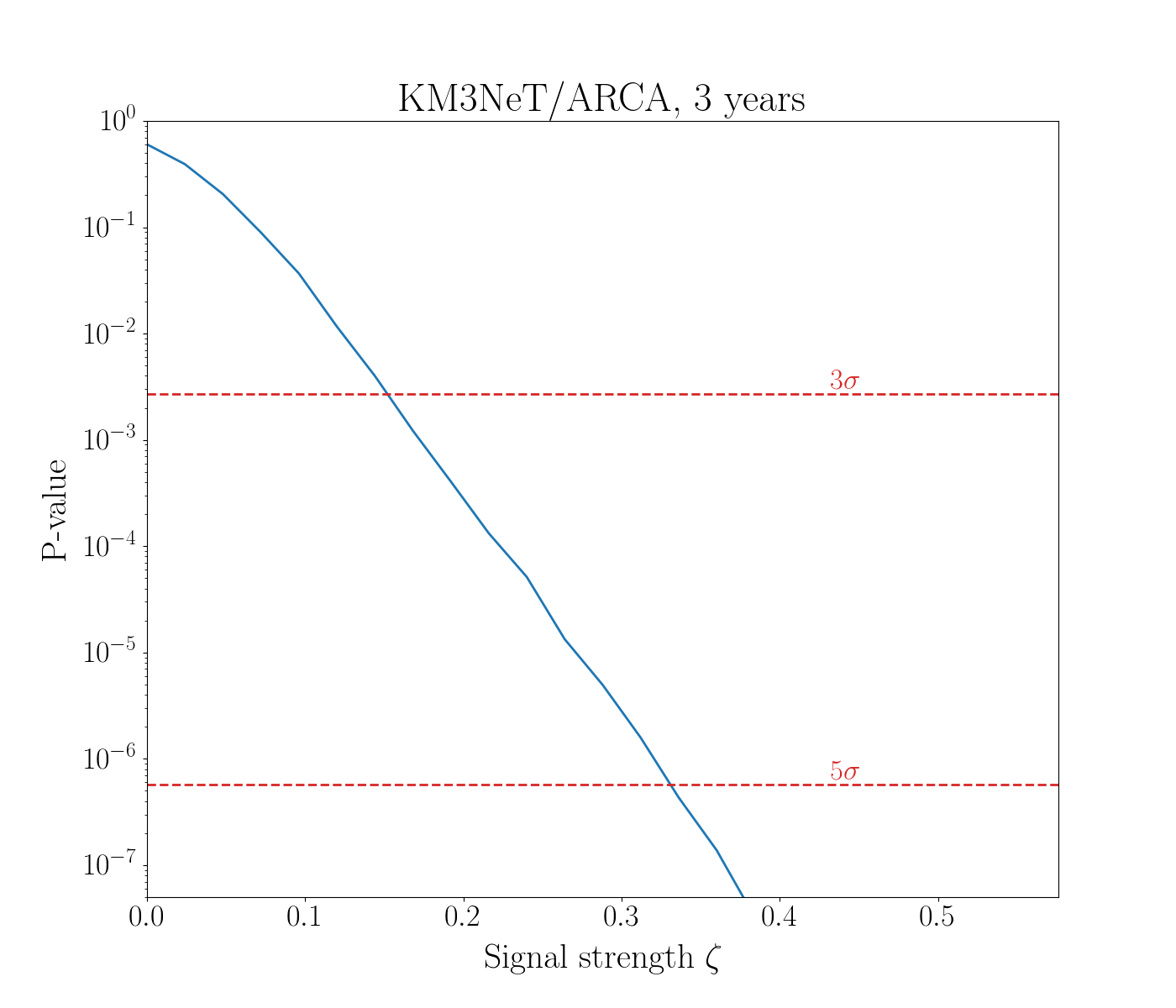} }}%
    \caption{Median confidence levels (a) and p-values of signal discovery (b) as function of the signal strength for a point-like source at sin($\delta$) = 0.1 using $\Phi_0 = 4 \times 10^{-9} \text{ GeV}^{-1} {\rm s}^{-1} {\rm cm}^{-2}$ and $\gamma = 2.0$ for three years of ARCA operation.}
    \label{fig:stattrack}%
\end{figure*}

The test statistic distributions are used to calculate two statistical quantities: the confidence level and the p-value. The median confidence level of rejecting a given signal strength is found by integrating a signal test statistic distribution from the median of the background-only distribution to infinity. This quantity can be used to calculate the neutrino flux for which the resulting test statistic would be higher than the median for background-only in 90\% of the cases, referred to as the sensitivity. The median two-sided p-value of discovering a given signal strength is calculated by integrating the background-only test statistic distribution from the median of a signal test statistic distribution to infinity. The p-values are used to determine the flux normalisation for which a 3$\sigma$ or 5$\sigma$ discovery could be claimed in 50\% of the pseudo-experiments. The median confidence levels and p-values as a function of the signal strength are shown in Figure \ref{fig:stattrack} for the example source.

\subsection{Results}

\begin{figure*}
\centering
\includegraphics[width=0.5\textwidth]{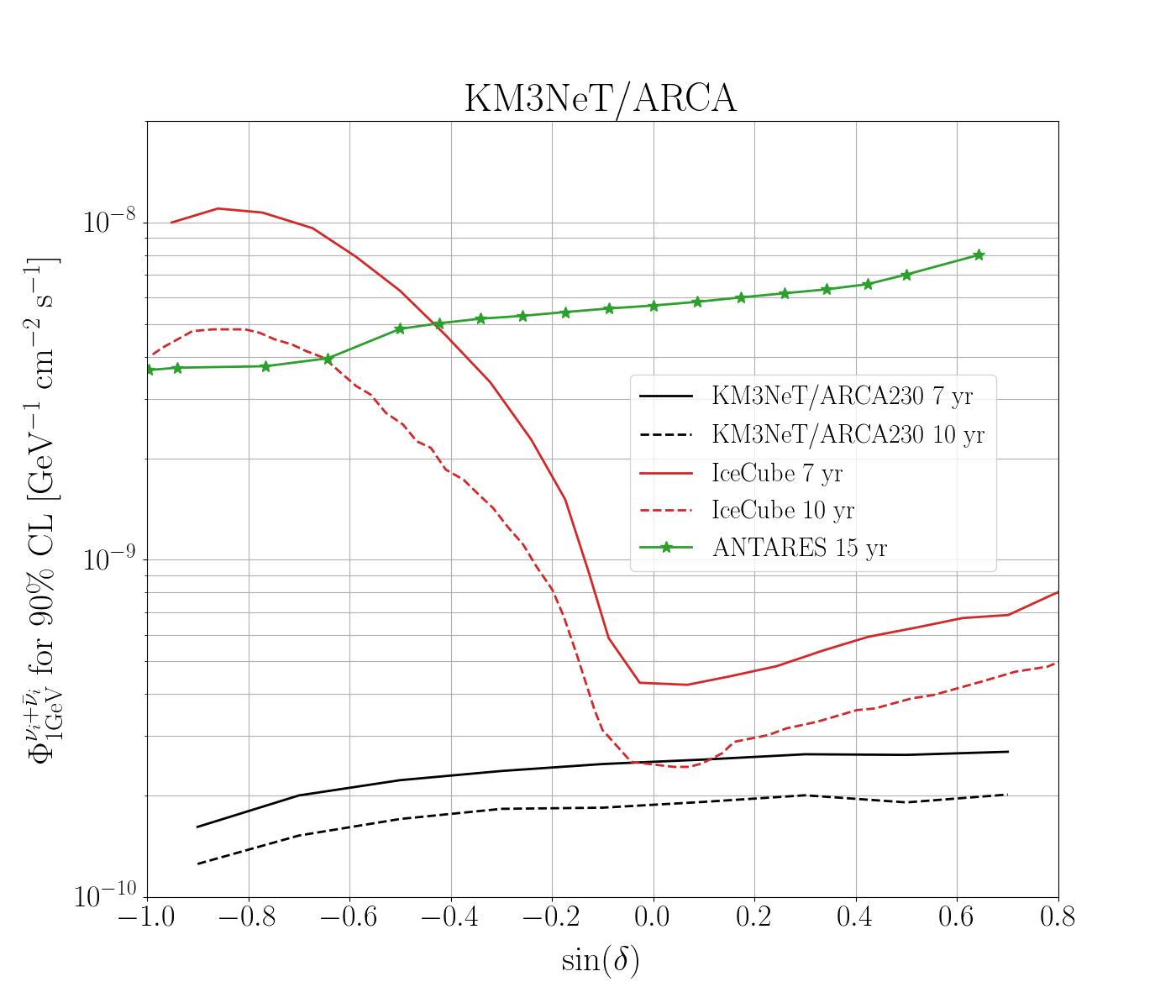}
\caption{ARCA point-like source sensitivity as a function of $\sin(\delta)$ for $\gamma=2.0$ (black curves). The results are compared with 15 years of ANTARES (green curve) \cite{alves2023antares} and 7 and 10 years of IceCube (red curves) \cite{aartsen2017all,aartsen2020time}.}
\label{fig:sensigamma2}
\end{figure*}

The point-like source sensitivity for a spectral index of $\gamma=2.0$ is given in Figure \ref{fig:sensigamma2} for 7 and 10 years of ARCA operation. The ARCA sensitivity is compared with the sensitivity of 15 years of ANTARES \cite{alves2023antares} and 7 and 10 years of IceCube \cite{aartsen2017all,aartsen2020time}. The sensitivity of ARCA surpasses that of IceCube in the Southern Sky ($\sin(\delta) < 0$) due to visibility. This is the proportion of time a source is below the horizon, favouring ARCA in the Northern Hemisphere to study the Southern Sky. In the Northern Sky ($\sin(\delta) > 0$), the enhanced sensitivity of ARCA is attributed to its improved angular resolution in comparison to IceCube. The recent Galactic Plane observation by the IceCube Collaboration \cite{icecube2023observation} puts ARCA in an excellent position to confirm this observation and to distinguish point-like source contributions from diffuse emission. 

\begin{figure*}
    \centering
    \subfloat[\centering  ]{{\includegraphics[width=0.45\textwidth]{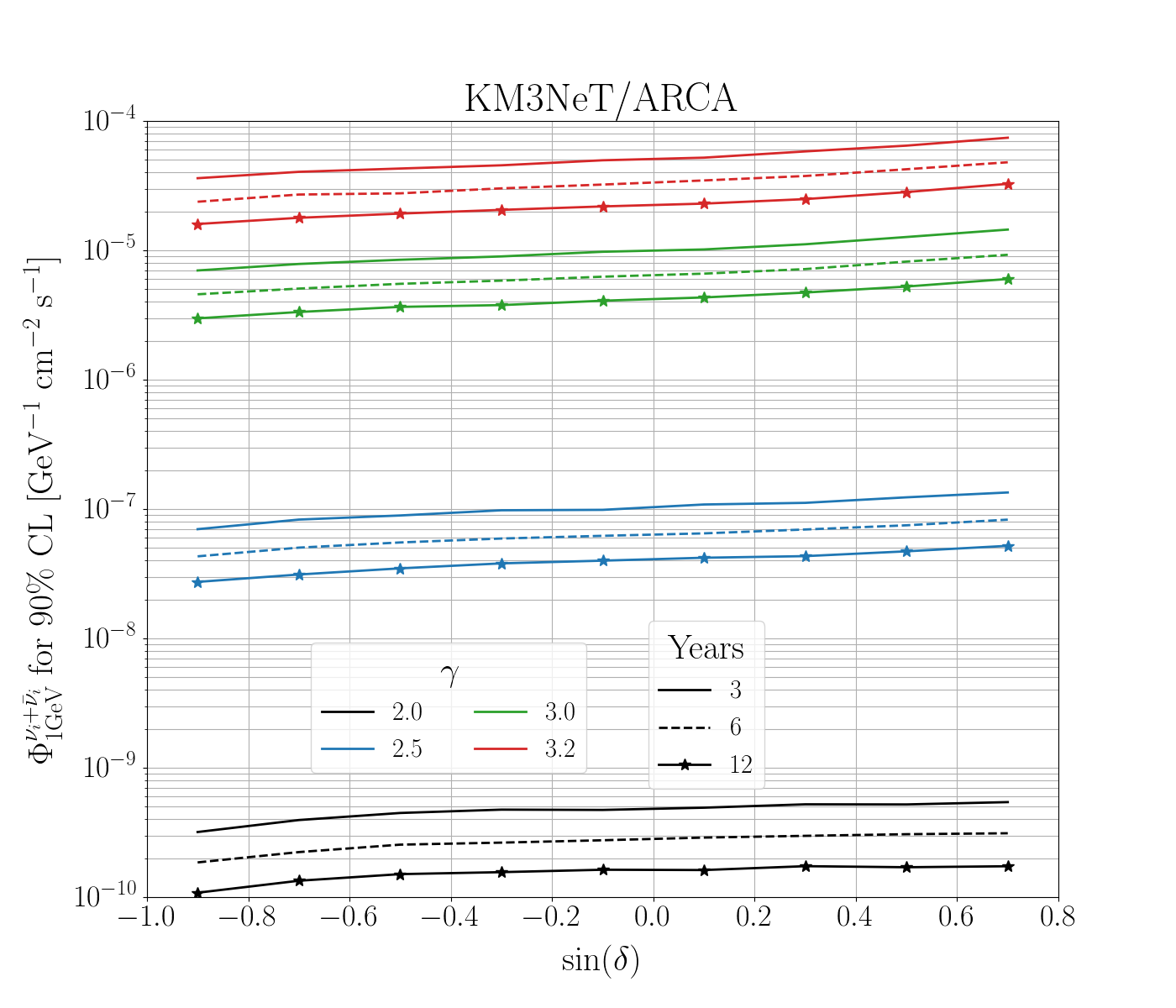}}}%
    \qquad
    \subfloat[\centering  ]{{\includegraphics[width=0.45\textwidth]{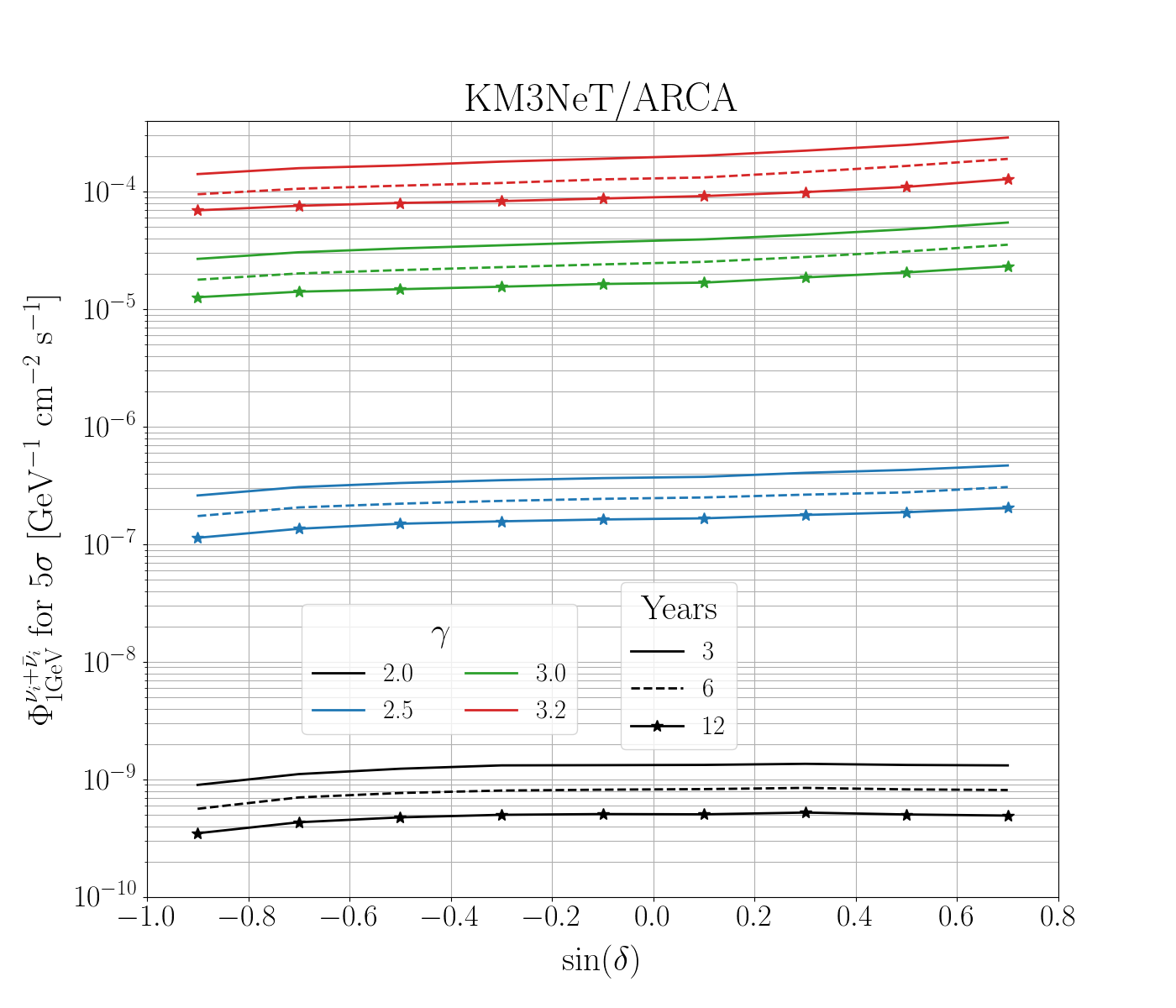} }}%
    \caption{ARCA Sensitivity (a) and 5$\sigma$ discovery flux (b) for different spectral indices and periods of data collection.}
    \label{fig:sensiallgamma_disc}%
\end{figure*}

The point-like source sensitivity for different spectral indices is given in Figure \ref{fig:sensiallgamma_disc}a. The $5\sigma$ discovery flux for point-like sources is given in Figure \ref{fig:sensiallgamma_disc}b for different spectral indices. The significant variation of flux normalisation across different spectral indices is due to the single power law of the flux model as shown in Equation \ref{eq:flux_point}.

%% file: systematics.tex
\section{Systematic uncertainties}

The current knowledge of systematic uncertainties was used to study the effects on the event reconstruction performances \cite{adrian2016letter}. These effects were used to modify the detector response functions of the analysis in order to determine the effect on the sensitivity and discovery potential of ARCA for point-like neutrino sources. The absorption and scattering length of light in the seawater at the ARCA site has been measured with 10\% accuracy \cite{riccobene2007deep}. Previous studies varied the water properties in the light simulation by $\pm$10\% to determine the influence on the direction and energy reconstruction. These effects were used to modify the detector response functions resulting in a $\pm$5-10\% variation of the sensitivity and discovery potential of the point-like analysis.

%% file: conclusions.tex
\section{Conclusion}

The ARCA detector is under construction and the first scientific results are being produced \cite{muller2023point,vasilis2023diffuse,francesco2023galactic}. The angular resolution of the detector for track-like and shower-like events offers the possibility of discovering new point-like neutrino sources. The sensitivity and discovery potential for point-like sources have improved by 5-15\% compared to previous results, thanks to improvements in event reconstruction performances, neutrino event selection, and the inclusion of shower events. ARCA has the capability to utilise both channels for investigating the Southern Sky, where many neutrino sources are expected along the Galactic Plane. This is particulary relevant in light of the recent discovery by the IceCube Collaboration of a diffuse neutrino emission along the Galactic Plane. ARCA will contribute to disentangling possible point-source contributions from the diffuse Galactic component.